\documentclass[12pt]{article}
\pdfoutput=1
\usepackage{color}
\usepackage{amsmath, amssymb}

\usepackage{mathrsfs}

\usepackage{ae} 
\usepackage[T1]{fontenc}
\usepackage[ansinew]{inputenc}
\usepackage{amsmath}
\usepackage{amssymb}
\usepackage{graphicx}
\usepackage{color}
\definecolor{darkblue}{cmyk}{0.9,0.9,0,0}
\usepackage[colorlinks=true,linkcolor=darkblue,citecolor=darkblue,urlcolor=darkblue]{hyperref}

\usepackage{array}

\newcommand{\beq}{\begin{equation}}
\newcommand{\eeq}{\end{equation}}
\newcommand\beqa{\begin{eqnarray}}
\newcommand\eeqa{\end{eqnarray}}
\newcommand\bea{\begin{array}}
\newcommand\eea{\end{array}}

\def\XXint#1#2#3{{\setbox0=\hbox{$#1{#2#3}{\int}$}
\vcenter{\hbox{$#2#3$}}\kern-.5\wd0}}

\newcommand{\nn}{\nonumber}

\newcommand{\COMMENT}[1]{}

\newcommand{\neqa}{\nonumber\end{eqnarray}}
\newcommand{\la}[1]{\label{#1}}

\newcommand{\<}{{\langle}}
\renewcommand{\>}{{\rangle}}

\newcommand{\re}{\relax{\rm I\kern-.18em R}}

\def\su2{{SU(2)}}

\def\[{\left[}
\def\]{\right]}

\def\({\left(}
\def\){\right)}
\def\[{\left[}
\def\]{\right]}

\def\<{\langle}
\def\>{\rangle}

\def\i2{\frac{i}{2}}

\usepackage{varioref}
\usepackage{makeidx}
\makeindex
\usepackage[english]{babel}

\begin{document}

\thispagestyle{empty}

\renewcommand{\thefootnote}{\fnsymbol{footnote}}
\setcounter{footnote}{0}
\setcounter{figure}{0}
\begin{center}
$$$$
{\Large\textbf{\mathversion{bold}
Tailoring Non-Compact Spin Chains
}\par}
\vspace{1.0cm}

Pedro Vieira$^{\mathcal A}$, Tianheng Wang$^{\mathcal{A}, \mathcal{B}}$
\\ \vspace{0.5cm}

\textit{$^{\mathcal A}$  Perimeter Institute for Theoretical Physics\\ Waterloo,
Ontario N2J 2W9, Canada} \\
\texttt{} \\
\vspace{.7mm}
\textit{$^{\mathcal B}$ Department of Physics and Astronomy \& Guelph-Waterloo Physics Institute,\\
University of Waterloo, Waterloo, Ontario N2L 3G1, Canada} \\
\texttt{} 
\vspace{0mm}

\par\vspace{.5cm}

\textbf{Abstract}
\end{center} 
\vspace{-.3cm}
We study three-point correlation functions of local operators in planar $\mathcal{N}=4$ SYM at weak coupling using integrability. 
We consider correlation functions involving two scalar BPS operators and an operator with spin, in the so called $SL(2)$ sector. At tree level we derive the corresponding structure constant for any such operator. We also conjecture its one loop correction.
To check our proposals we analyze the conformal partial wave decomposition of known four-point correlation functions of BPS operators. In perturbation theory, we extract from this decomposition sums of structure constants involving all primaries of a given spin and twist. On the other hand, in our integrable setup these sum rules are computed by summing over all solutions to the Bethe equations. A perfect match is found between the two approaches. 

\vspace*{\fill}

\setcounter{page}{1}
\renewcommand{\thefootnote}{\arabic{footnote}}
\setcounter{footnote}{0}

\newpage

\tableofcontents

\newpage 
\section{Introduction and Main Result}\la{intro}

\begin{figure}[t]
\centering
\includegraphics[scale=0.5]{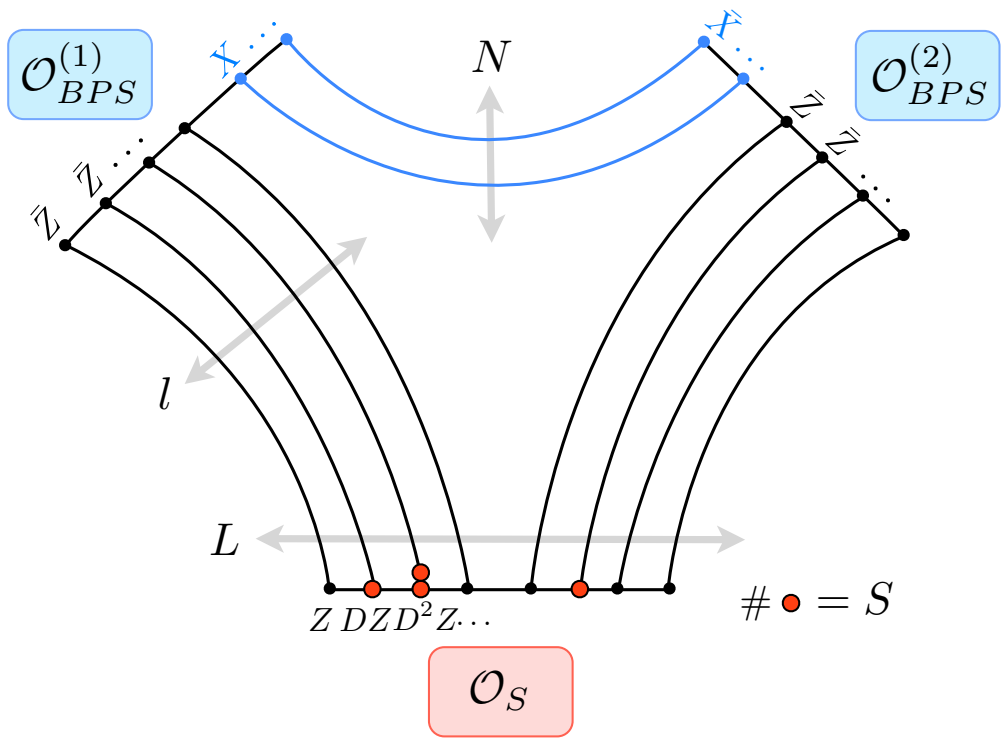}\vspace{-.5cm}
\caption{In this paper we consider a correlation function of two $SU(2)$ BPS operators and one non-protected $SL(2)$ primary operator of spin $S$, twist $L$ and dimension $\Delta$.  The BPS operators are given by the sum over all the positions of inserting some complex scalars $X$ or $\bar X$ in a sea of complex scalars $\bar Z$, see (\ref{BPSop}). The non-BPS spin-$S$ operator is more interesting and its form is governed by a non-trivial wave function, see (\ref{generalForm}). 
} \label{Fig3pt} 
\end{figure}

In this paper we study three-point correlation functions of local operators in planar $\mathcal{N}=4$ SYM at tree level and at one loop using the underlying integrable structure of the theory \cite{review}. We will study the structure constant $C^{\bullet \circ\circ}$ governing the correlation functions involving two scalar BPS operators and a spin $S$ operator in the so called $SL(2)$ sector as depicted in figure \ref{Fig3pt}. 
Our work generalizes some of the results in \cite{tailoringI,tailoringIV} from the compact $SU(2)$ case to the non-compact $SL(2)$ setup. 
We start by describing our setup and presenting our main result (\ref{mainResult}). 

We consider the correlation function
\beq
\<\mathcal{O}_{S}(x_1)\mathcal{O}_{BPS}^{(1)}(x_2)\mathcal{O}_{BPS}^{(2)}(x_3) \> =\frac{1}{N_c} \, \frac{C^{\bullet\circ\circ}}{x_{12}^{\Delta-S+2\sc{l}-L} x_{13}^{\Delta-S+L-2\sc{l}} x_{23}^{L+2N-(\Delta-S)}} \(\frac{x_{12}^+}{x_{12}^2}-\frac{x_{13}^+}{x_{13}^2}\)^S \la{3pt} \,,
\eeq
as depicted in figure \ref{Fig3pt}.  The structure constant $C^{\bullet\circ\circ}$ is physical once we normalize the two point functions, 
\beq
\<\mathcal{O}_{BPS}^{(1)}(x)\bar{\mathcal{O}}_{BPS}^{(1)}(0) \> =   \frac{1}{x^{2l+2N}}\,, \,\,\, \<\mathcal{O}_{BPS}^{(2)}(x)\bar{\mathcal{O}}_{BPS}^{(2)}(0) \> =   \frac{1}{x^{2L-2l+2N}} \,, \,\,\, \<\mathcal{O}_{S}(x)\bar{\mathcal{O}}_{S}(0) \> =\frac{(x^+)^S}{x^{\Delta+S}}  \, ,\la{2pt}
\eeq
where bar stands for complex conjugation.\footnote{To be more precise, this normalization condition does not fix the structure constant completely since we can always multiply any of the three operators by a phase. This does not affect the two point functions (\ref{2pt}) but changes the phase of $C^{\bullet\circ\circ}$ in (\ref{3pt}). Hence, by itself the structure constant is not a physical quantity but its absolute value is. 
In this paper we always use the freedom of tuning the phase of the external operators to set the structure constant in (\ref{3pt}) to be real. 
}

Each protected operators is taken to be in a $SU(2)$ sector and is therefore parametrized by two integers that indicate how many complex scalars it is made of, see figure \ref{3pt}.\footnote{We assume that $N\ge 1$ but the exact value of $N$ will be irrelevant for the most part. For example, the dependence of $N$ is trivial and factorizes in our main result (\ref{mainResult}). The reason why we do \textit{not} consider the $N=0$ case is that in that case the length of the non-BPS operator would be equal to the sum of the lengths of the two BPS operators. In other words, there would be no propagators on the top of figure \ref{Fig3pt} and this correlator would be \textit{extremal}. For extremal correlators we also need to take into account the operator mixing of the large operator with double traces which is an annoying complication. On the other hand, this same mixing is suppressed at large $N_c$ if the correlator is not extremal. This is why we want at least a small \textit{bridge} on top of figure \ref{Fig3pt}, i.e. $N\ge 1$. This same reasoning led to the $SU(2)$ setup of \cite{tailoringI} for scalar operators. \la{Npositive}} For example,  
\beq
\mathcal{O}_{BPS}^{(1)}(x) \propto  \sum_{1\le n_1 < \dots < n_{N} \le \sc{l}+N}\,\,\,\,\,\, {\rm Tr}\Big(\overbrace{\underbrace{\bar{Z}\dots \bar{Z}}_{n_1-1}\bar{X}\bar{Z}\dots \bar{Z}}^{n_2-1}\bar{X}\bar{Z} \dots \Big)(x) \,. \la{BPSop}
\eeq
The operator $\mathcal{O}_{BPS}^{(2)}(x)$ is given by a similar expression with $\sc l \to L- \sc l$ and with the complex scalar $\bar {X}$ replaced by its conjugate $X$.

The non-BPS spin S operator is more interesting and its form is governed by a non-trivial wave function  
\beq
\mathcal{O}_{S}(x) =  \sum_{1\le n_1 \le n_2 \le \dots \le n_{S} \le L} \psi(n_1,\dots,n_S) \, \mathcal{O}_{n_1,\dots,n_S}(x)
\la{generalForm}
\eeq
where $\mathcal{O}_{n_1,\dots,n_S}(x)$ stands for an operator with $L$ scalars and derivatives at positions $n_1$, $n_2$, etc. We also include some conventional $1/m!$ numerical coefficients if $m$  derivatives act on the same scalar field. That is, 
\beq
\mathcal{O}_{n_1,\dots,n_S}(x)=\[ \prod_{j=1}^L \frac{1}{m_j!} \]  \, {\rm Tr}
\Big(\underbrace{\overbrace{{Z}\dots {Z}}^{n_1-1}D_+{Z}\dots {Z}}_{n_2-1}D_+{Z} \dots \Big)(x)  \la{On1n2}
\eeq
where $D_+$ is a covariant derivative in a light-like direction and $m_j$ stands for the number of derivatives acting on the $j$-th scalar $Z$. For example $\mathcal{O}_{1,2,4}=  {\rm Tr}
((D_+Z) (D_+ Z) Z (D_+ Z) Z  Z \dots )$ and $\mathcal{O}_{2,2,4}= \frac{1}{2!} {\rm Tr}
(Z (D_+^2 Z) Z (D_+ Z) Z Z  \dots )$ etc.

For a generic twist $L$ and spin $S$, there are several possible primary operators (\ref{generalForm}) corresponding to different possible wave functions $\psi$. 
These wave functions are  found by requiring that the states (\ref{generalForm}) diagonalize the quantum corrected $\mathcal{N}=4$ dilatation operator. 

At tree level we have a large degeneracy of several primary operators with the same classical dimension $L+S$. As we turn on the coupling, these dimensions acquire quantum corrections and this degeneracy is lifted. Since the degeneracy is lifted already at one loop, the one loop eigenstates are enough to parametrize the states at any order in perturbation theory. Both the leading order eigenstates and their first loop corrections are described in detail in Section \ref{DrAndGs}.

For now their precise form is not important. It suffices to know that each primary operator is parametrized by a set of real numbers $\{u_1,\dots,u_S\}$ called Bethe roots. These Bethe roots are constrained by a set of so called Bethe ansatz equations that arise once we impose periodicity for the wave function in (\ref{generalForm}) and that take the form \cite{Braun:1999te,MatthiasSmatrix,Minahan:2002ve,Beisert:2003yb}
\begin{eqnarray}
e^{i p(u_j) L} \prod_{k\neq j}^S \mathcal{S}(u_j,u_k)=1 \,, \qquad \prod_{j=1}^S e^{i p(u_j) } =1 \,. \la{BAE}
\end{eqnarray}
In this expression the momentum $p(u)$ and the S-matrix $\mathcal{S}(u,v)$ are best parametrized using the so called Zhukovsky variables $x(u)$ given by 
\beq
x(u) \equiv \frac{u+\sqrt{u^2-4g^2}}{2g} \,, \qquad x_j^{\pm} \equiv x(u_j\pm i/2) \,,
\eeq
where the coupling 
\beq
g^2 = \frac{g^2_{YM} N_c}{16\pi^2} \,.
\eeq
In terms of these,  
\begin{eqnarray}
e^{i p(u_j)} = \frac{x^+_j}{x^-_j} \, , \qquad \mathcal{S}(u_j,u_k)= \frac{u_j -u_k + i}{u_j - u_k -i}\left[\frac{1-\frac{1}{x^-_j x^+_k}}{1-\frac{1}{x^+_j x^-_k}}\right]^2 \sigma^2(u_j,u_k)  \,. \la{pS} 
\end{eqnarray}
The BES dressing phase $\sigma^2(u,v)$ \cite{Beisert:2006ez} is irrelevant and can be set to $1$ throughout this paper since it first deviates from $1$ at four loops which is way beyond the scope of this work. 

The different solutions to the Bethe equations (\ref{BAE}) are in one to one correspondence with the different possible primaries (\ref{generalForm}). The quantum corrected dimension $\Delta$ of the operator $\mathcal{O}_S$ -- which appears in the exponents in (\ref{3pt}) and (\ref{2pt}) -- is simply given by 
\beq
\Delta=L+S+\gamma \, , \qquad \gamma= 2ig\sum_{j=1}^S \( \frac{1}{x^+_j}- \frac{1}{x^-_j}\) \,. \la{energy}
\eeq
It is not hard to count the number of solutions to (\ref{BAE}) for a given spin $S$ and twist $L$. 

For example, for twist $L=2$ there is a single solution to (\ref{BAE}) for even $S$ and no solution for odd spin. This means that for twist $2$ there is actually no degeneracy at all. This makes the study of these operators and their correlation functions considerably simpler and, indeed, these are the operators that are studied in greater depth in the literature. 

For twist $L \ge 3$ we start having several solutions. For example, for twist $L=4$ we have
\beq
\begin{array}{l}\text{number of solution to (\ref{BAE}) for}\\
\text{$L=4$ and spin $S=2,3,4,5,6,7,8,9,\dots$} \end{array}=\{2,2,5,4,8,8,13,12,\dots\}
\eeq

The structure constant $C^{\bullet\circ\circ}$ in (\ref{3pt}) depends explicitly on the integers $N$, $L$, $\sc{l}$, $S$, on the coupling $g$ and on the set of Bethe roots $\{u_1,\dots,u_S\}$ solving (\ref{BAE}). The purpose of this paper is to study this quantity at tree level and one loop in the planar limit. We derived this quantity at tree level and proposed a conjecture for its value at one loop. In total, our main result reads
\begin{eqnarray}
C^{\bullet\circ\circ} =\frac{\sqrt{L(\sc l+N)(L-\sc l+N)}}{\sqrt{
\left(\!\!\begin{array}{c}\sc l+N\\N\end{array}\!\!\right)\left(\!\!\begin{array}{c}L-\sc l+N\\N\end{array}\!\!\right)}} \left(1-\frac{\gamma }{2}\right)\frac{ \mathcal A_{\sc l}  }{\mathcal B }\,+O(g^4) \,.\label{mainResult}
\end{eqnarray}
The result (\ref{mainResult}) is, not surprisingly, strikingly similar to the analogue $SU(2)$ result -- see equation (24) in \cite{tailoringIV} and Appendix \ref{su2} for a detailed comparison. The contribution $\mathcal{B}$ depends uniquely on the non-BPS operator and is given by a simple determinant,
\begin{eqnarray}\la{calB}
\mathcal B =\sqrt{ {\frac{1}{\prod\limits_{j=1}^S \frac{\partial p(u_j)}{\partial u_j} } \left| \det_{1\le j,k \le S}\frac{\partial}{\partial u_j}\left[L p(u_k)+\frac{1}{i}\sum_{l\neq k}^S\log \mathcal S(u_k, u_l)\right]\right|}}\, . 
\end{eqnarray}
The most interesting contribution is $\mathcal{A}_{\sc{l}}$ which depends explicitly on the three-point function setup through the integer $\sc{l}$, see figure \ref{Fig3pt}. This contribution can be written as a sum over all possible ways of splitting the Bethe roots $\{u_j\}$ into two partitions, 
\begin{eqnarray}
\mathcal A_{\sc l} = \frac{\sum\limits_{{\color{red}\alpha}\cup{\color{blue} \bar\alpha} = \{u_j\}} (-1)^{|{\color{red} \alpha}|} \prod\limits_{\color{blue} u_j\in \bar\alpha} e^{-ip({\color{blue} u_j})l} \prod\limits_{\substack{{\color{red} u_j\in\alpha}\\ {\color{blue} u_k\in\bar\alpha}}} \mathfrak f({\color{red} u_j},{\color{blue} u_k})}{\sqrt{\prod\limits_{j\neq k}\mathfrak f(u_j,u_k)}\prod\limits_{j} \(e^{-ip(u_j)}-1\)}\, . \la{calA}
\end{eqnarray}
Finally, in this expression, the function $\mathfrak{f}(u,v)$ reads 
\begin{eqnarray}\la{frakf}
\mathfrak{f}(u,v) = \frac{u-v+i}{u-v}\left(1+g^2\frac{i (u-v-i)}{(u^2+1/4) (v^2+1/4)}+O(g^4)\right). 
\end{eqnarray}
The main difference to the SU(2) result -- reviewed in appendix \ref{su2} -- is the form of this function $\mathfrak{f}(u,v)$.\footnote{At tree level, for $g=0$, the object (\ref{calA}) has a nice determinant representation as can be easily derived following \cite{Ivan,Ivan2}; would be nice to investigate whether there is also such a representation incorporating the first quantum correction.}

This concludes the discussion of our main result for the structure constants in (\ref{3pt}). 

It would be very interesting to generalize this computation to more general correlation functions, where more than one operator has spin, see e.g. \cite{volodya} and \cite{penedones} for interesting works in this direction. It would also be instructive to study interesting limits of our conjecture (\ref{mainResult}) such as large spin limits, in the spirit of \cite{tailoringIII} and \cite{fernando}. Other very interesting limits to play with would be those where the integer spin is analytically continued to complex values and taken to extreme values such as $S\to -1$ where all loop constraints  \cite{joaoMiguel} might guide us figuring out the next quantum corrections to (\ref{mainResult}).
Our work, when combined with \cite{tailoringI,tailoringIV} and \cite{JoaoThiago}, provides valuable hints about the structure of correlation functions of generic local operators in planar $\mathcal{N}=4$ SYM theory. It would be very interesting to combine all these results into a single description of very general correlators.

We present some non-trivial checks of (\ref{mainResult}) against available perturbative data in section \ref{checksSec}. In section \ref{DrAndGs} we present the derivation/motivation of our conjecture. We also present some speculative remarks in that section. Sections \ref{checksSec} and \ref{DrAndGs} can be read independently of each other. Additional details are presented in the appendices.

\section{Conjecture versus Data} \la{checksSec}
In this section we compare our prediction (\ref{mainResult}) with the results in the literature obtained by direct perturbative computations. We find perfect agreement with all the available data. 

\subsection{Twist 2 operators}\la{twist2}
We start by studying the simplest possible case in figure \ref{Fig3pt} where the non-BPS operator has minimal twist $L=2$ (and $l=1$). 
As is well known, for even spin there is a single primary operator of the form (\ref{generalForm}) with twist two and for odd spin there is no primary operator at all with this twist. Indeed, there is a single solution to the Bethe equations (\ref{BAE}) for $L=2$ and $S$ even and there are no solutions for $L=2$ and $S$ odd. Furthermore, for $L=2$, the Bethe roots $u_j = u_j^{(0)}+ g^2 u_j^{(1)}+ \dots $ are also particularly simple to find. To leading order, they are given by the zeros of a Hahn polynomial \cite{Korchemsky:1994um},\beq
\prod_{j=1}^S(u-u_j^{(0)}) = c_S \,_3 F_2\left(-S,S+1,\frac{1}{2}- iu ;1,1;1\right) \,, \qquad c_S=\frac{(S!)^2}{2^S (-1)^{S/2}(2S-1)!!}   \,.
\eeq
With \verb"Mathematica", the roots of the Hahn polynomials for each spin $S$ can be computed with arbitrary precision. Once the leading order position of the Bethe roots is found, the quantum corrections $u_j^{(1)}$ are computed by linearizing the Bethe equations around this solution. This can again be done with arbitrarily high precision. Finally, the Bethe roots are plugged into  (\ref{mainResult}). It turns out that the final result for the (square of the) structure constant (\ref{mainResult}) up to one loop can be expressed in terms of rational numbers. For example, for the first few spins we find, 
\beqa
\(C_{\text{twist-2}}^{\bullet\circ\circ} \)^2 &=& \frac{1}{3}-4g^2+ \mathcal{O}(g^4) \qquad\qquad \quad\text{for $S=2$} \nn \\
\(C_{\text{twist-2}}^{\bullet\circ\circ} \)^2 &=&\frac{1}{35} - \frac{205}{441} g^2+ \mathcal{O}(g^4) \qquad\quad\; \text{for $S=4$} \la{Plefka1}\\
\(C_{\text{twist-2}}^{\bullet\circ\circ} \)^2 &=& \frac{1}{462} - \frac{1106}{27225} g^2+ \mathcal{O}(g^4) \qquad \text{for $S=6$} \nn
\eeqa
and so on. 
These are the integrability based predictions for the correlator in figure \ref{Fig3pt} for $L=2$, $l=1$ and generic $N$ (in this simple case the $N$ dependence cancels out). 

At tree level this structure constant was first computed by Dolan and Osborn in \cite{Dolan:2000ut}  by analyzing the operator product expansion of four BPS operators.
Since there is a single primary with twist $L=2$ and two units of $R$-charge its contribution to the OPE is particularly simple to single out. This analysis was generalized to one loop in \cite{Arutyunov:2003ad}. The results of \cite{Dolan:2000ut} and \cite{Arutyunov:2003ad} read 
\begin{eqnarray}
\( C_{\text{twist-2}}^{\bullet\circ\circ} \)^2 = \frac{2 (S!)^2}{(2S)!} \[1- 4 g^2 \sum_{j=1}^S \(\frac{1}{j^2}+ \frac{2}{j}  \sum_{k=S+1}^{2S}\frac{1}{k}\) +\mathcal O(g^4)\]\,. \la{Plefka}
\end{eqnarray}
which perfectly agree with our predictions (\ref{Plefka1}). Recently, this same three-point function was also computed directly in perturbation theory in \cite{Plefka:2012rd}.   

Recently, the computation of four point functions of BPS operators in the so called ${\bf 20'}$ representation was revived due the discovery of a hidden permutation symmetry \cite{KorchemskyHidden} which completely determines the correlation function integrand up to remarkably high loop orders \cite{MoreHidden}. 
Eden managed to OPE decompose these results thus predicting the value of the structure constant in (\ref{Plefka}) up to three loops for arbitrary spin $S$ \cite{Eden:2012rr}. Any proposal for the higher loop corrections to the integrability result (\ref{mainResult}) ought to reproduce this formidable amount of data. Hopefully this can be used as a powerful guiding principle when looking for such corrections.

\subsection{Higher twist operators}\la{highertwist}
We shall now repeat the previous analysis for operators with larger twist. One important difference is that for larger twist $L$, the spin $S$ is not enough to uniquely specify the primary operator. Instead, there are several primary operators with the same spin and these are in one to one correspondence with the several solutions to the Bethe equations (\ref{BAE}). 
For each primary operator we can predict the corresponding structure constant in (\ref{3pt}) by simply plugging the corresponding solution to the Bethe equations $\{u_1,\dots,u_S\}$ into (\ref{mainResult}). For example, for $L=4$ and spin $S=4$ there are five solutions to (\ref{BAE}). One such solution is given by 
\begin{eqnarray}
&& \quad u_1=  -1.7535503703709001299 - 3.4849705606400700999 \,g^2\, , \nonumber\\
&& \quad u_2= -0.8988670992551022430 - 2.8259960143725117275 \,g^2\, ,\nonumber\\
&& \quad u_3= -0.3916751257167127515 - 2.0137468513733101546 \,g^2\, ,\nonumber\\
&& \quad u_4= -0.0607423439092896214 + 0.0219639498301160823 \,g^2\, .\la{firstExample}
\end{eqnarray}
(It is trivial to arbitrarily increase the precision of these roots as needed.) We can now plug these Bethe roots into (\ref{mainResult}) to obtain structure constant in (\ref{3pt}) for the corresponding primary operator. 

We should also specify $N$ and $l$ which parametrize the BPS operators, see figure \ref{Fig3pt}. For $L=4$ the most symmetric setup is the one where the BPS operator is evenly \text{split} in two, that is for $l=2$. The $N$ dependence is not so interesting since it factors out in (\ref{mainResult}) but we do want it to be positive, see footnote \ref{Npositive}. Hence, for concreteness let us consider the simplest possible case corresponding to $N=1$ and $l=2$.  

Then, we find, for the Bethe roots (\ref{firstExample}), the following prediction:
\beq
\left(C^{\bullet\circ\circ}\right)^2=0.01037344398340248963 - 0.1624417241056073032 \,g^2 \,. \la{prediction}
\eeq
We would now like to check this result against a direct field theory computation. 

More precisely, the tree level result in (\ref{prediction}) does not need to be checked since it is derived in section \ref{treeDerivaion} and hence it is definitely correct. However we would like to check the one loop correction which is a conjecture. Unfortunately, there is no direct perturbative computation of structure constants of higher twist operators. The work \cite{Plefka:2012rd} considered twist $2$ operators only. It would be very interesting to generalize \cite{Plefka:2012rd} to arbitrary twist and check our predictions such as  (\ref{prediction}) for arbitrary twist $L$ and spin $S$ primaries. 

In the meantime, to check out conjectures, we consider a small detour. We will see that while we can not match the values of individual structure constants such as (\ref{prediction}) we can easily check particular sums of structure constants involving all possible primaries of a given spin and twist.

\subsection{Sum Rules and the OPE Decomposition}\la{twist4Main}

One way of obtaining structure constants which does not rely on computing directly three point functions in perturbation theory is by analyzing four point correlation functions. In this approach one decomposes these objects in conformal partial waves and reads off dimensions and structure constants of the operators flowing in this decomposition \cite{Dolan:2000ut}. 
In perturbation theory we expand around the point $g=0$ where there is a huge degeneracy. Hence, the conformal partial wave decomposition -- in its most obvious form -- yields particular sums of structure constants for primary operators with the same twist and spin.  This approach is particularly powerful because from a single four-point function we can extract infinitely many such sums for  infinitely many operators that flow in the OPE. 

With this motivation in mind, we now split our discussion into two parts. We start with 
a discussion on the computation of such sums from the integrability point of view. Afterwards we match these against explicit OPE decompositions of four-point functions available in the literature. 
 
The sums $\mathcal{P}^{(n,m)}_S$ that will arise in the OPE can be concisely described with the generating function\footnote{
We added a subscript ${\bf u}$ to the anomalous dimension (\ref{energy}) and to the structure constant (\ref{mainResult}) to emphasize that they depend on the particular solution to the Bethe equations. 
In (\ref{defSums}), $y$ is simply a bookkeeping parameter used to define the generating function. Since the anomalous dimensions $\gamma_{\bf u}=\mathcal{O}(g^2)$, once expanded in perturbation theory, the $y$-dependence of the right hand side of (\ref{defSums}) matches that in the left hand side such that the sums $\mathcal{P}^{(n,m)}_S$ are properly defined. Their (even more) explicit expressions are given in appendix \ref{explicitSums}.}
\beq
\sum_{n=0}^\infty g^{2n}  \sum_{m=0}^n y^m \,\mathcal{P}^{(n,m)}_S \equiv \sum_{\begin{array}{c} \text{\scriptsize solutions ${\bf u}=\{u_1,\dots,u_S\}$ to} \\ \text{\scriptsize BAE (\ref{BAE}) with fixed $S$ and $L$} \end{array}} \!\!\!\!\! \!\!\!\!\! \(C_{\bf u}^{\bullet\circ\circ} \)^2 \exp\(\gamma_{\bf u}\, y \)\,. \la{defSums}
\eeq
The sums $\mathcal P_S^{(n,m)}$ only depend on the integers $S$, $L$ as well as $l$ and $N$, recall figure \ref{Fig3pt}. 
In appendix \ref{explicitSums}, they are explicitly written down in terms of the (loop corrections to the) structure constants and anomalous dimensions. 
The sums $\mathcal{P}_S^{(n,m)}$ with $m=0,1,\dots,n$ appear in the decomposition of a four point function at $n$ loops. 

With our one loop prediction (\ref{mainResult}) we can already predict an infinite subset of these sums. More precisely we can predict the values of $\mathcal{P}_S^{(n,n)}$ and $\mathcal{P}_S^{(n,n-1)}$ for any $n$. With a two loop conjecture for the structure constant we could also predict $\mathcal{P}_S^{(n,n-2)}$ and so on. 

It turns out that the sums $\mathcal{P}_S^{(n,m)}$ are somehow much nicer than the individual structure constants like (\ref{prediction}) as we now illustrate. For that we consider, as above, the case $L=4$, $S=4$, $l=2$, $N=1$ and compute $\mathcal{P}_S^{(0,0)}$, $\mathcal{P}_S^{(1,0)}$ and $\mathcal{P}_S^{(1,1)}$. For $L=S=4$  there are five solutions to Bethe equations (\ref{BAE}); one of them yields the structure constant (\ref{prediction}). We now find all five solutions, plug them in (\ref{energy}) and (\ref{mainResult}) and add them up as in (\ref{defSums}). We find a nice surprise.
We obtain
\beq
\!\!\!\!\!\!\!\!\sum_\text{five sols}\!\!\! \(C_{\bf u}^{\bullet\circ\circ} \)^2 e^{ \gamma_{\bf u} y } =  0.{\bf 253968}\,{ 253968}\,{\bf 253968} - 2.{\bf 419753086}\,{ 419753086} \,g^2 +   2.{\bf 444444444}  \,g^2 y\nonumber + \dots \,, 
\eeq
which is a clear rational number (we adjusted the precision of the several terms to highlight the periodic nature of the digits characteristic of rational numbers). That is, we get 
\beq
\sum_\text{five sols}\!\!\! \(C_{\bf u}^{\bullet\circ\circ} \)^2 e^{ \gamma_{\bf u} y } =  \frac{16}{63}- \frac{196}{81}g^2+\frac{22}{9} g^2 y + \dots \, . \nn
\eeq
In the same way we can predict several more sums for different values of $S$ (keeping the same twist $L=4$ and also $N=1$ and $l=2$ as above). For example, for $n\leq 3$ we find Table \ref{Pnm}.
\begin{table}[t]
\renewcommand{\arraystretch}{1.5}
\caption{Values of $\mathcal P^{(n,m)}_S$ for $L=4$, $l=2$ and $N=1$.}\label{Pnm}
\centering
\begin{tabular}{|c|ccccccc|}
 \firsthline
Spin-$S$ & $0$ &$ 2$ & $4$  &$6$ & $8$ & $10$ &$\dots $ \\ \hline  
$\mathcal P^{(0,0)}$ & $4$ &$\frac{7}{5}$  & $\frac{16}{63}$ & $ \frac{29}{858}$ & $\frac{46}{12155}$ & $\frac{67}{176358}$  & $\dots$ \\ \hline 
$\mathcal P^{(1,1)}$ & $0$ &$ 8$ &  $\frac{22}{9}$ & $\frac{892}{2145}$ & $\frac{419}{7735}$  & $\frac{4307}{712215}$ & $ \dots$ \\ \hline
$\mathcal P^{(1,0)}$ & $0$ &$ -8$ &  $-\frac{196}{81}$ & $ -\frac{1873528}{4601025}$ & $-\frac{12573551}{239320900}$  & $-\frac{1686168467}{289857260700}$ & $ \dots$ \\ \hline
$\mathcal P^{(2,2)}$ & $0$ & $24$ & $\frac{112}{9}$ & $\frac{171679}{64350}$ & $\frac{975976}{2436525}$ & $\frac{88719937}{1794781800}$ & $\dots$ \\ \hline
$\mathcal P^{(2,1)}$ & $0$ & $-64$ & $ -\frac{827}{27}$ & $ -\frac{3495054427}{552123000} $ &  $-\frac{209769902257}{226158250500}$  & $-\frac{10945855305643}{97392039595200} $ & $\dots $ \\ \hline
$\mathcal P^{(3,3)}$ & $0$ & $\frac{160}{3}$ & $\frac{3688}{81}$ & $\frac{17359409}{1447875}$ & $\frac{58218457}{28426125}$ & $\frac{470324411813}{1696068801000}$ & $\dots$ \\ \hline
$\mathcal P^{(3,2)}$ & $0$ & $-256$ & $-\frac{16054}{81}$ & $-\frac{254442194}{51122500}$ & $-\frac{781946179598053}{94986465210000}$ & $-\frac{17879207919405953}{16434906681690000}$ & $\dots$\\
\lasthline
 \end{tabular}
 \end{table}

All the very non-trivial looking rational numbers in table \ref{Pnm} can be matched against perturbative data by OPE decomposing appropriate four-point functions as we now explain. 

Consider the correlation function of four BPS operators
\begin{eqnarray}
\< \,\text{Tr} (ZZX)(x_1)  \,\text{Tr} (ZZ\bar X)(x_2)\, \text{Tr}(\bar Z\bar Z Y)(x_3)\,  \text{Tr} (\bar Z\bar Z\bar Y)(x_4)\> \equiv \frac{1}{x_{12}^{6} x_{34}^{6}}\,\mathcal G(z,\bar z) \la{4pt}
\end{eqnarray}
where $z$ and $\bar z$ are the usual cross-ratios 
\beq
\frac{(x_1-x_2)^2 (x_3-x_4)^2}{(x_1-x_3)^2 (x_2-x_4)^2} = z \bar z\, ,\quad\quad \frac{(x_1-x_4)^2 (x_2-x_3)^2}{(x_1-x_3)^2 (x_2-x_4)^2} = (1-z)(1-\bar z)\, , \la{cr}
\eeq
which 
behave as $z,\bar z \to 0$ in the OPE limit $x_{1}\to x_2$ which we will be considering. The OPE expansion is a double expansion in $z$ and $\bar z$ where powers of $z$ are the conformal spins (dimension \textit{plus} spin) of the exchanged operators whereas powers of $\bar z$ measure their twists (dimension \textit{minus} spin), see \cite{Dolan:2000ut} for more details. 
 
Now, the OPE of $\text{Tr} (ZZX)(x_1)$ and $\text{Tr} (ZZ\bar X)(x_2)$ produces operators with $4$ units of $R$-charge in the $Z$ direction. Hence, the operators with the smallest possible twist in this OPE are of the form of (\ref{generalForm}) with twist $L=4$.  Therefore, the contribution of the leading twist operators to $\mathcal G(z,\bar z)$ is completely governed by the sums $\mathcal{P}_S^{(n,m)}$ introduced above. 
This contribution is given by the leading terms as $\bar z\to 0$ as we now review. More precisely we have, 
\beq
\mathcal G(z,\bar z)=\bar z^{L/2} f(z,\tau) + \mathcal{O}(\bar z^{L/2+1}) 
\eeq
At order $n$ in perturbation theory $f(z,\tau)$ is a polynomial in $\tau \equiv \frac{1}{2} \log(z\bar z)$ of degree $n$ given by 
\beq
f(z,\tau) = \sum_{n=0}^\infty g^{2n} \sum_{m=0}^n \sum_{S=0}^\infty \mathcal{P}_S^{(n,m)} f_S^{(m)}(z,\tau) \la{OPEexpansion}
\eeq
where the functions $f_S^{(m)}(z,\tau)$ are fixed by conformal symmetry and take the form 
\beq\la{fS}
f_S^{(m)}(z,\tau) =\left. z^{\frac{L+2S}{2}} \frac{\partial^m}{\partial \gamma^m} \( e^{\tau \gamma}\, {}_2 F_1\(\frac{L+2S+\gamma}{2},\frac{L+2S+\gamma}{2};L+2S+\gamma;z\) \)\right|_{\gamma=0}\,.
\eeq
This is easily derived from the $\bar z\to 0$ expansion of the standard conformal blocks, see e.g. \cite{Dolan:2000ut}. 
These functions admit a regular expansion at small $z$, 
\beq
f^{(m)}_S(z,\tau) = {\color{magenta} z^{\frac{L+2 S}{2}}}  \left(  \tau^m  + 
\[ \tfrac{\partial^m}{\partial \gamma^m}   \tfrac{e^{\gamma \tau}(L+2 S+\gamma)}{4}\]_{\gamma=0}   {\color{magenta} z} + \[\tfrac{\partial^m}{\partial \gamma^m} \tfrac{e^{\gamma  \tau } (L+2 S+\gamma ) (L+2 S+\gamma +2)^2}{32 (L+2 S+\gamma +1)}\]_{\gamma=0}    {\color{magenta} z^2} + \mathcal{O}({\color{magenta} z^3}) \right) \,.\nn
\eeq
The larger $S$ is, the more suppressed $f_S^{(m)}$ are in the OPE limit.

Finally, since  many of the $\mathcal{P}_S^{(n,m)}$'s in (\ref{OPEexpansion}) can be predicted from (\ref{mainResult}), we can predict a great deal about the behaviour of the four point function in the OPE limit $z,\bar z\to0$. Using the values in table \ref{Pnm} in (\ref{OPEexpansion}) we get the prediction\footnote{At each loop order we can predict the two highest powers of $\tau$ from the knowledge of the tree level and one loop structure constants. Starting at two loops, however, there are also lower powers of $\tau$ which can only be predicted once the two loop and higher structure constants are computed.
} 
\begin{eqnarray}
\!\!\!\!f(z,\tau)\!\!\!&=&\!\!\!  g^0 \Big[{ \color{blue}{4} z^2 +{4} z^3 + {5} z^4 + {6} z^5 + {7} z^6 +\cdots}\Big] \nonumber\\
\!\!\!\!\!\!\!\!\!&+&\!\!\!  g^2\Big[ {\color{blue}  \tau\left(\! 8 z^4 + 16 z^5 + \frac{74}{3} z^6  +\cdots\!\right)-   \left(\!8 z^4 + 14 z^5 + \frac{179}{9} z^6+\cdots\!\right)}\Big] \nonumber\\
 \!\!\!\!\!\!\!\!\!&+&\!\!\! g^4\Big[  {\color{blue} \tau^2 \left(\!{ 24} z^4 + { 48} z^5 +{\frac{712}{9}} z^6 \cdots\!\right)-   \tau \left(\! {64} z^4 + {116} z^5 + {\frac{1619}{9}} z^6 +\cdots\!\right)}+\mathcal{O}(\tau^0) \Big] \nn \\
\!\!\!\!\!\! \!\!\!&+& \!\!\! g^6 \Big[{\color{magenta} \tau^3 \left(\! \frac{160}{3} z^4 + \frac{320}{3} z^5 + \frac{15688}{81} z^6 +\cdots \right) - \tau^2 \left(256 z^4 + 472 z^5 +\frac{214480}{729} z^6+\cdots\right)} +\mathcal O(\tau) \Big] \nonumber\\
\!\!\!\!\!\! \!\!\!&+& \!\!\! \mathcal O(g^8)\,.\la{voila}
\end{eqnarray}

We can now compare the prediction (\ref{voila}) against a direct perturbative computation of the four-point correlation function (\ref{4pt}). This quantity was computed up to two loops in \cite{Arutyunov:2003ad}. Expanding their result in the OPE limit we find {(see appendix \ref{detailsT4} for more details)}
\begin{eqnarray}
\!\!\!\!f(z,\tau)_{\text{ref \cite{Arutyunov:2003ad}}}\!\!\!&=&\!\!\!  g^0 \Big[{\color{blue} {4} z^2 +{4} z^3 + {5} z^4 + {6} z^5 + {7} z^6 +\cdots}\Big] \nonumber\\
\!\!\!\!\!\!\!\!\!&+&\!\!\!  g^2\Big[ {\color{blue}  \tau\left(\! 8 z^4 + 16 z^5 + \frac{74}{3} z^6  +\cdots\!\right)-   \left(\!8 z^4 + 14 z^5 + \frac{179}{9} z^6+\cdots\!\right)}\Big] \nonumber\\
 \!\!\!\!\!\!\!\!\!&+&\!\!\! g^4\Big[{ \color{blue}  \tau^2 \left(\!{ 24} z^4 + { 48} z^5 +{ \frac{712}{9}} z^6 \cdots\!\right)-   \tau \left(\! { 64} z^4 + { 116} z^5 + { \frac{1619}{9}} z^6 +\cdots\!\right)}\nn \\
 \!\!\!\!\!\!\!\!\!&&\qquad  {\color{magenta}+\left(8(7+6\zeta_3) z^4 + 96(1+\zeta_3) z^5 +\left(\frac{7805}{54}+148 \zeta_3\right)z^6\cdots\right)}\Big] \nn \\
\!\!\!\!\!\! \!\!\!&+&\!\!\!\mathcal{O}(g^6)\,. \la{pertvoila}
\end{eqnarray}

We now discuss the comparison between  (\ref{voila}) and  (\ref{pertvoila}). 
First, and most importantly, we note that the terms that are captured by both expressions perfectly match! These are the two leading powers of $\tau$ up to two loops --  shown in blue in the first three lines of (\ref{voila}) and (\ref{pertvoila}). The remaining terms in these expressions -- coloured in magenta -- are also very interesting as we now discuss. 

From the terms in magenta in the perturbative computation (\ref{pertvoila}) we can now read off the sums $\mathcal{P}^{(2,0)}_S$, see table \ref{P20}.
\begin{table}[t]
\renewcommand{\arraystretch}{1.5}
\caption{$\mathcal P_S^{(2,0)}$ from the OPE for $L=4$, $l=2$, $N=1$.} \label{P20}
\centering
\begin{tabular}{|c|cccccc|}
\firsthline
Spin-$S$ & $0$ &$ 2$ & $4$  &$6$ & $8$  &$\dots $ \\ \hline
$\mathcal P^{(2,0)}_S$ & $0$ & $56 + 48\zeta_3$ &  $\frac{12397}{486} + \frac{44}{3} \zeta_3$ & $\frac{24650493920837}{4737215340000 }+ \frac{1784}{715}\zeta_3$ & $\frac{786038480222357}{1036642410440000}+\frac{2514}{7735}\zeta_3$ & $\dots$ \\
\lasthline
\end{tabular}
\end{table}
All these numbers should be matched against any candidate for the next quantum correction to (\ref{mainResult}). For example, one could try to cook up an educated guess for the two loop correction to (\ref{mainResult}) and use this data to constrain the potential of this guess. We played a bit with these ideas but we were not imaginative enough to come up with the right ansatz thus far. 

The terms in magenta in the integrability expression (\ref{voila}) are also interesting. They provide predictions to the next loop corrections (3 loops and higher) to the four-point correlation function (\ref{4pt}) which might be useful in constraining or simplifying the perturbative computation of this object. 

We repeated this comparative analysis for several other cases. In appendix \ref{twist3}, for example, we present the analogue results for the sum rules $\mathcal{P}^{(n,m)}_S$ for twist $L=3$ (and $l=1$) and also for twist $L=6$ (and $l=2$). 

Finally, we should point out that, we could also consider slightly more general sums involving a product of two different structure constants. That is, we could construct 
\beq
\sum_{\begin{array}{c} \text{\scriptsize solutions ${\bf u}=\{u_1,\dots,u_S\}$ to} \\ \text{\scriptsize BAE (\ref{BAE}) with fixed $S$ and $L$} \end{array}} \!\!\!\!\! \!\!\!\!\! C_{\bf u}^{\bullet\circ\circ}  \widetilde{C}_{\bf u}^{\bullet\circ\circ}   \exp\(\gamma_{\bf u} x \) \equiv  \sum_{n=0}^\infty g^{2n}  \sum_{m=0}^n y^m \,\widetilde{\mathcal{P}}^{(n,m)}_S \la{defSumsNew}
\eeq
which would govern the OPE behaviour of a less symmetric correlation function (compared to (\ref{4pt}) where all external operators have the same size). For example, from the OPE analysis of a correlation function such as 
\begin{eqnarray}
\< \,\text{Tr}(Z X)(x_1)\,  \text{Tr} ( Z Z  Z \bar X)(x_2)\, \text{Tr} (\bar Z\bar ZY)(x_3)  \,\text{Tr} (\bar Z\bar Z\bar Y)(x_4)\, \> \equiv \frac{1}{x_{12}^{6} x_{34}^{6}}\frac{x^2_{14}}{x^2_{24}}\,\tilde{\mathcal G}(z,\bar z) \la{4ptNew} \,,
\end{eqnarray}
we would be able to read off the sums (\ref{defSumsNew}) where ${C}_{\bf u}^{\bullet\circ\circ} $ would correspond to (\ref{mainResult}) with $L=4$, $N=1$ and $l=1$ while $\widetilde {C}_{\bf u}^{\bullet\circ\circ} $ would be given also by (\ref{mainResult}) but for $L=4$, $N=1$ and $l=2$. 

In table \ref{tildePnm}, in appendix \ref{twist3}, we present the integrability predictions for the sum rules (\ref{defSumsNew}) for this case. From these sums, following the same kind of analysis as above, we could predict that $\tilde{\mathcal G}(z,\bar z) = \bar z^{2} \tilde f(z,\tau) + \mathcal{O}(\bar z^3)$ where $\tilde f(z,\tau)$ is given in (\ref{f2433}) in appendix \ref{twist3}.\footnote{Note that for external operators of different dimensions the conformal blocks need to be slightly modified in a straightforward way (see for example, \cite{Dolan:2000ut}) and this leads to some obvious minor modifications to the functions (\ref{fS}) arising in the expansion of $f(z,\tau)$.} Chicherin and Sokatchev kindly shared with us their unpublished two loop result for (a generalization of) the four-point correlation function (\ref{4ptNew}) \cite{emeryToAppear}. We OPE expanded their proposal finding perfect agreement with our predictions for $\tilde f(z,\tau)$. From their two loop results one can also extract the first two loop data for the asymmetric sum rules (\ref{defSumsNew}). 

We should emphasize again that all these detours -- involving constructing and checking sum rules for structure constants rather than the structure constants themselves -- steam from the absence of any perturbative results for three-point functions involving operators of generic twist. It would be very interesting to develop further the perturbative side of this story and directly check our predictions such as (\ref{prediction}) against a direct perturbative computation of the three point function (\ref{3pt}). 

Alternatively -- and given that these sum rules seem to be simpler than the individual terms in the sum -- it would be interesting to develop an integrability based approach for computing the sums directly.

\section{Derivation (tree level) and educated guess (1 loop) } \la{DrAndGs}
In this section we explain how (\ref{mainResult}) arises from an integrability based approach. At tree level, we derive this result; at one loop it is an educated guess whose motivation we will present. 

\subsection{The (ultra-local) wavefunctions for the operator $\mathcal{O}_S$} \la{ultraLocal}
One important ingredient in the three-point function (\ref{3pt}) is the form of the non-BPS operator (\ref{generalForm}). Since we are interested in the \textit{one loop} structure constants we need the non-BPS operators at $\mathcal{O}(g^2)$. These operators diagonalize the \textit{two loop} planar dilatation operator of $\mathcal{N}=4$ SYM. 

Strictly speaking, the form of an operator is not a very physical quantity since we can always change it by performing field redefinitions. In other words, we can always apply similarity transformations to the dilatation operator. Still, we can adopt a particular scheme. Physical quantities  such as the structure constants and the operator dimensions will not depend of that choice. 

In the literature there are two representations of the dilatation operator, related by a similarity transformation. One was worked out by Eden and Staudacher in \cite{Eden:2006rx} and the other by Zwiebel in \cite{Zwiebel:2008gr}. Each has its own advantages and drawbacks. The representation of \cite{Eden:2006rx} is very simple and explicit however it was worked out only for operators with spin $S=1$, $2$ or $3$. The representation of \cite{Zwiebel:2008gr} can be applied for operators of any spin however it is considerably harder to manipulate.\footnote{In the representation \cite{Zwiebel:2008gr} the Hamiltonian is written in terms of a bilinear of supercharges $Q$. The full Hamiltonian acts inside the $SL(2)$ sub-sector (\ref{generalForm}) as expected but each individual supercharge does not. Hence, to deal with this representation, in intermediate states we need to consider more general operators, with fermions and so on. For large spins, this makes the use of this representation very cumbersome, even using a computer.} We checked explicitly that for $S=1,2$ or $3$ the two Hamiltonians are related by a similarity transformation. We will mostly use \cite{Eden:2006rx}; some comments on the comparison with \cite{Zwiebel:2008gr} are presented in appendix \ref{D2}.

For spin $S=1,2$ and $3$ the eigenvectors of the dilatation operator of \cite{Eden:2006rx} take the form (\ref{generalForm}) with the wave functions
\beqa
 \psi(n_1) &=&   \phi_1 \la{waveFunctions}\\
 \psi(n_1,n_2) &=&  \phi_{12}+\mathcal{S}_{12} \phi_{21} 
\nn\\ 
  \psi(n_1,n_2,n_3)&=& \phi_{123}+\mathcal S_{12}\phi_{213}+\mathcal S_{23}\phi_{132}+\mathcal S_{23}\mathcal S_{13}\phi_{312}+\mathcal S_{12}\mathcal S_{13}\phi_{231}+\mathcal S_{12}\mathcal S_{13}\mathcal S_{23}\phi_{321}
\nn
\eeqa
We now explain in detail the meaning of these symbols. 
First note 
that we can also change the overall normalization of the wave functions such that the two-point functions satisfy (\ref{2pt}). 
The S-matrices $\mathcal{S}_{ab}=\mathcal{S}(u_a,u_b)$ in (\ref{waveFunctions}) appeared already in the introduction. As mentioned there, the Bethe rapidities $u_a$ are a particularly nice parametrization of the momenta $p_a=p(u_a)$ which are quantized according to the Bethe equations (\ref{BAE}). Finally, we have the plane waves
\beq
\phi_{a_1\dots a_S}=e^{i p_{a_1} n_1+\dots +i p_{a_S} n_S} \(1+g^2 \delta_{a_1\dots a_S}\)
\eeq
where $\delta_{a_1\dots a_S}$ are the so called contact terms or fudge factors introduced in \cite{MatthiasSmatrix}. They are zero if the particles are well separated. Let us postpone their discussion for now. 

Clearly, these wave functions have a transparent physical meaning. They describe a set of particles which scatter among themselves in a factorized way. Their form is typical of integrable models with local interactions and is said to be of \textit{Bethe ansatz form}. The Bethe equations (\ref{BAE}) are nothing but the periodicity condition for these wave functions. The generalization to more particles is then straightforward: we simply add up $S!$ plane waves decorated by the appropriate products of S-matrices. 

It is also clear from this physical picture that a \textit{Bethe ansatz} might need to be slightly improved if the interactions have some finite range. When the particles are within the interaction range the wave functions should be corrected. In $\mathcal{N}=4$ SYM the planar dilatation operator can be thought of as a local Hamiltonian whose range increases by one unit at each order in perturbation theory. The contact terms $\delta_{a_1\dots a_S}$ precisely take into account the finite range nature of the interactions and correct the wave function for nearby particles. For a single particle $\delta_{a}=0$. For two particles the most generic contact term that we might expect to encounter at two loop order would take the form
\begin{eqnarray}
\delta_{ab} = \left\{
\begin{array}{ll}
0 & \quad \text{for } n_2-n_1>1 \, ,   \\
\mathbb{C}_{\bullet\bullet}(p_a,p_b) & \quad \text{for } n_2-n_1=1 \,,\\ 
\mathbb{C}_{\substack{\bullet\vspace{-.6mm}\\ \bullet}}(p_a,p_b) & \quad \text{for } n_2-n_1=0 \, .\end{array}
\right. \la{C2}
\end{eqnarray}
while for three particles the most general form of contact terms would be  
\begin{eqnarray}
\delta_{abc} = \left\{
\begin{array}{ll}
0 & \quad \text{for } n_2-n_1>1\text{ and } n_3-n_2>1 \, , \\
\mathbb{C}_{\bullet\bullet}(p_a,p_b) & \quad \text{for } n_2-n_1=1 \text{ and } n_3-n_2>1\, , \\
\mathbb{C}_{\bullet\bullet}(p_b,p_c) & \quad \text{for } n_2-n_1>1 \text{ and } n_3-n_2=1\, , \\
\mathbb{C}_{\substack{\bullet\vspace{-.6 mm}\\ \bullet}}(p_a,p_b) & \quad \text{for } n_2-n_1=0 \text{ and } n_3-n_2>1\,,\\
\mathbb{C}_{\substack{\bullet\vspace{-.6 mm}\\ \bullet}}(p_b,p_c) & \quad \text{for } n_2-n_1>1 \text{ and } n_3-n_2=0\,,\\
\mathbb{C}_{\substack{\bullet\;\, \vspace{-.6 mm}\\ \bullet\bullet}}(p_a,p_b,p_c) & \quad \text{for }n_2-n_1=0 \text{ and } n_3-n_2=1\,,\\
\mathbb{C}_{\substack{\,\;\bullet\vspace{-.6 mm}\\ \bullet\bullet}}(p_a,p_b,p_c) & \quad \text{for } n_2-n_1=1 \text{ and } n_3-n_2=0\, ,\\
\mathbb{C}_{\bullet\bullet\bullet}(p_a,p_b,p_c) & \quad \text{for } n_2-n_1=1 \text{ and }n_3-n_2=1\,, \\
\mathbb{C}_{\substack{\bullet\vspace{-.6 mm}\\ \bullet\vspace{-.6 mm} \\ \bullet}}(p_a,p_b,p_c) & \quad \text{for } n_2-n_1=0 \text{ and }n_3-n_2=0\, .\end{array}
\right. \la{C3}
\end{eqnarray}
Again, so far, everything we wrote is very generic and would be roughly the same for any integrable model with such finite (but short) range interactions. This formalism was dubbed as asymptotic Bethe ansatz in \cite{Sutherland,MatthiasSmatrix}.

We will now review a very special feature of this particular $SL(2)$ spin chain which we dub as \textit{ultra-local} nature of the contact terms. It turns out that we only have non-zero contact terms at one loop when the particles are right on top of each other. There is no contact terms when they are next to each other, 
\beq
\mathbb{C}_{\bullet\bullet}(p_a,p_b) =0
\eeq
and 
\beq
\mathbb{C}_{\bullet\bullet\bullet}(p_a,p_b,p_c) =0\,, \qquad \mathbb{C}_{\substack{\bullet\;\, \vspace{-.6 mm}\\ \bullet\bullet}}(p_a,p_b,p_c) = \mathbb{C}_{\substack{\bullet\vspace{-.6mm}\\ \bullet}}(p_a,p_b)  \, , \qquad  \mathbb{C}_{\substack{\,\;\bullet\vspace{-.6 mm}\\ \bullet\bullet}}(p_a,p_b,p_c) =\mathbb{C}_{\substack{\bullet\vspace{-.6mm}\\ \bullet}}(p_b,p_c) \la{3local}
\eeq
That is, up to three particles we actually only need two contact terms: $\mathbb{C}_{\substack{\bullet\vspace{-.6mm}\\ \bullet}}(p_a,p_b) $ and $\mathbb{C}_{\substack{\bullet\vspace{-.6 mm}\\ \bullet\vspace{-.6 mm} \\ \bullet}}(p_a,p_b,p_c) $. This was also pointed out in \cite{Eden:2006rx}. 
The precise form of the two relevant contact terms is written in the appendix \ref{D1}, see (\ref{ES1}) and (\ref{ESC3}). 

At this point it is very natural to assume that this ultra-local simplification generalizes in the most obvious way to four and more particles as well. This is the most important outcome of all this discussion and we will make use of this later when arguing for (\ref{mainResult}).
Unfortunately, since the Hamiltonian in \cite{Eden:2006rx} was only written up to $S=3$ it is not possible to straightforwardly check this conjecture for a few more cases with $S=4,5$ etc.

Finally, we found that the norm of the states (\ref{waveFunctions}) are given by the typical Gaudin's norm. This was checked at tree level already in \cite{tailoringI} but here we checked it for the quantum corrected states (\ref{waveFunctions}) as well. We found that 
\beq
\sum_{1\le n_1 \le \dots \le n_S \le L} |\psi(n_1,\dots,n_S) |^2 =\left| \det\limits_{1\le j,k \le S} \frac{\partial}{\partial p_j} \(L p_k+\frac{1}{i} \sum\limits_{l\neq k} \log \mathcal{S}(u_k,u_j)\) \right|\la{Gaudin}
\eeq
In the structure constant this factor should appear in the denominator since we should normalize the non-BPS operator $\mathcal{O}_S$ as in (\ref{2pt}). This explains the factor $1/\mathcal{B}$ in (\ref{mainResult}). In the next sections we discuss the factor $\mathcal{A}_l$, the most interesting part of the full result.

\subsection{Tree-level derivation}\la{treeDerivaion} 
At tree level we can know everything about all the three states in (\ref{3pt}) and all we need to do is Wick contract them to compute the structure constant (\ref{mainResult}). We now sketch the derivation, following \cite{tailoringI} closely (see in particular section 3.1 in \cite{tailoringI}).

To evaluate (\ref{3pt}) at tree level we split the bottom state into two spin chains of length $l$ and $L-l$ as depicted in figure \ref{Fig3pt}. In the left and right subchains we can have any number $M$ and $S-M$ of derivatives, respectively.\footnote{This constitutes one important difference compared to the $SU(2)$ case in \cite{tailoringI}. There, conservation of $R$-charge fixed the number of excitations in each subchain.} 
Very explicitly, by plugging the states (\ref{generalForm}) and (\ref{BPSop}) into (\ref{3pt}) and performing all Wick contractions we end up with 
\beqa
&&\!\!\!\!\!\!\!\!\!\! \<\mathcal{O}_{S}(x_1)\mathcal{O}_{BPS}^{(1)}(x_2)\mathcal{O}_{BPS}^{(2)}(x_3) \> =\frac{1}{N_c}
\frac{\mathcal{N}}{x_{23}^{2N}}  \sum_{M=0}^S\left[ \sum_{\alpha \cup \bar \alpha=\{u_j\} \text{ with } |\alpha|=M} \prod_{\bar a \in \bar \alpha} \( e^{il p_{\bar a}} \prod_{a \in \alpha} \mathcal{S}(u_a,u_{\bar a}) \)  \!\! \la{complicated} \right.\\
&&\qquad\qquad\qquad \qquad \qquad \times   \(\sum_{1\le n_1 \le \dots \le n_M \le l} \, \psi_{\alpha}(n_1,\dots,n_M) 
\prod_{j=1}^{l} \[\frac{1}{m_j!}\(\!\frac{\partial}{\partial{x_1^+}}\!\)^{\!\!m_j}\!\!\! \frac{1}{x_{12}^2}\]\)  \nn\\
&&\left.\qquad\qquad\qquad \qquad \qquad \times  \( \sum_{1 \le n_{M+1} \le \dots \le n_{S} \le L-l}  \psi_{\bar \alpha}(n_{M+1},\dots,n_S) \prod_{j=1}^{L-l}\[ \frac{1}{m_j!}\(\!\frac{\partial}{\partial{x_1^+}}\!\)^{\!\!m_{j}}\!\!\! \frac{1}{x_{13}^2}\] \) \right]\nn
\eeqa
where $\mathcal{N}$ is a simple factor due to the normalization of each of the three operators; for example, it contains the Gaudin norm (\ref{Gaudin}) discussed in the previous section.\footnote{The relation between the two point function of spin operators and the Gaudin norm is most easily established by employing a basis where Wick contractions and spin chain scalar products are equivalent, see e.g. footnote 12 in \cite{back}. 
} 
This expression can be dramatically simplified as we now explain. 
First note that for each value of $M$ we get a space dependence proportional to\footnote{The factors of $m_j$ in (\ref{On1n2}) neatly drop out when taking the derivatives of the propagators in (\ref{complicated}).} $$\frac{1}{x_{23}^{2N} x_{12}^{2l}\, x_{13}^{2(L-l)}} \(\frac{x_{12}^+}{x_{12}^2}\)^M \!\!\(\frac{x_{13}^+}{x_{13}^2}\)^{\!\!S-M} \,.$$
From conformal symmetry, we know that at the end of the day the sum over $M$ in (\ref{complicated}) must be proportional to the last factor in (\ref{3pt}) 
\beq
\(\frac{x_{12}^+}{x_{12}^2}-\frac{x_{13}^+}{x_{13}^2}\)^S =\sum_{M=0}^S (-1)^{S-M} \binom{S}{M}  \(\frac{x_{12}^+}{x_{12}^2}\)^{M} \(\frac{x_{13}^+}{x_{13}^2}\)^{S-M} 
\eeq
Hence, we can simply consider the $M=S$ case since all other terms must yield the same result up to a simple combinatorial factor. The $M=S$ term in (\ref{complicated}) simplifies enormously since $\bar\alpha$ is the empty set. 
Therefore
\beq
 \<\mathcal{O}_{S}(x_1)\mathcal{O}_{BPS}^{(1)}(x_2)\mathcal{O}_{BPS}^{(2)}(x_3) \> =\frac{1}{N_c} \, \frac{\(\frac{x_{12}^+}{x_{12}^2}-\frac{x_{13}^+}{x_{13}^2}\)^S }{x_{12}^{2l}\, x_{13}^{2(L-l)} x_{23}^{2N}}  \,  \mathcal{N}  \(\sum_{1\le n_1 \le \dots \le n_S \le l} \, \psi(n_1,\dots,n_M) \) \,.
\eeq
Furthermore, the sum in the parentheses is nothing but the scalar product between an $SL(2)$ off-shell Bethe state and a vacuum descendent. Those scalar products can be computed straightforwardly by  changing a few signs in the $SU(2)$ result -- see appendix A.2 of \cite{tailoringI} -- and yield the factor $\mathcal{A}_l$ in (\ref{mainResult}). This concludes the sketch of the tree level derivation.

\subsection{One-loop conjecture}

In the previous section we derived (\ref{mainResult}) at tree level. At one loop the structure constant receives $g^2$ corrections due to two different effects. 
\begin{itemize}
\item On the one hand the wave functions in (\ref{generalForm}) get corrected. This includes corrections to the S-matrices of the excitations as well as the inclusion of the contact terms discussed in section \ref{ultraLocal}. 
Unfortunately we do not have a solid description of the quantum corrected states for $S\ge 4$. 
\item On the other hand, at one loop, we need to add loops to the tree Wick contractions discussed in the previous section. This second effect can be taken into account by inserting a splitting operator acting on the legs in figure \ref{Fig3pt} at the splitting points as described in \cite{BeforeTailoring,Alday:2005nd}. Such operators are well understood for three-point functions involving scalars \cite{BeforeTailoring,Alday:2005nd}. Unfortunately, for operators involving derivatives they are not known in full generality. It should be possible to generalize the results of \cite{Plefka:2012rd,Alday:2005nd} to eliminate this gap.
\end{itemize}
Given our ignorance about either type of corrections we  have to resort to some guesswork to motivate (\ref{mainResult}).

Our main assumption is that the form of the result is a minor deformation of the tree level result. That is we assume that 
\begin{eqnarray}
C^{\bullet\circ\circ} =\frac{\sqrt{L(\sc l+N)(L-\sc l+N)}}{\sqrt{
\left(\!\!\begin{array}{c}\sc l+N\\N\end{array}\!\!\right)\left(\!\!\begin{array}{c}L-\sc l+N\\N\end{array}\!\!\right)}} \times \,\texttt{simple prefactor} \times \frac{ \mathcal A_{\sc l}  }{\mathcal B }\,+O(g^4) \,,\label{mainResult2}
\end{eqnarray}
where 
\begin{eqnarray}\la{calB2}
\mathcal B =\sqrt{ {\frac{1}{\prod\limits_{j=1}^S \frac{\partial p(u_j)}{\partial u_j} } \left|\det_{1\le j,k \le S}\frac{\partial}{\partial u_j}\left[L p(u_k)+\frac{1}{i}\sum_{l\neq k}^S\log \mathcal S(u_k, u_l)\right]\right|}}\, , 
\end{eqnarray}
and 
\begin{eqnarray}
\mathcal A_{\sc l} = \frac{\sum\limits_{{\color{red}\alpha}\cup{\color{blue} \bar\alpha} = \{u_j\}} (-1)^{|{\color{red} \alpha}|} \prod\limits_{\color{blue} u_j\in \bar\alpha} e^{-ip({\color{blue} u_j})l} \prod\limits_{\substack{{\color{red} u_j\in\alpha}\\ {\color{blue} u_k\in\bar\alpha}}} \mathfrak f({\color{red} u_j},{\color{blue} u_k})}{\sqrt{\prod\limits_{j\neq k}\mathfrak f(u_j,u_k)}\prod\limits_{j} \(e^{-ip(u_j)}-1\)}\, . \la{calA2}
\end{eqnarray}
In this ansatz $\mathcal{S}(u,v)$ is the loop corrected $S$-matrix which is known. So the only unfixed ingredients in our guess are the quantum corrections to the prefactor \beq
\texttt{simple prefactor}=1+\mathcal{O}(g^2) \nn
\eeq and to the function 
\beq\mathfrak{f}(u,v)=\frac{u-v+i}{u-v}+\mathcal{O}(g^2)\,.\la{ftree}\eeq 
It is very important to point out that the ansatz (\ref{mainResult2}) is not a random guess. On the contrary, in the computation of a similar structure constant -- but with the non-BPS operator made out of scalars (instead of derivatives) -- the quantum corrected result took exactly this form \cite{tailoringIV}. This is the main motivation for (\ref{mainResult2}). 

In the $SU(2)$ scalar case the quantum corrected structure constants were actually derived rigorously in  \cite{tailoringIV} and,  
in particular, the outcome of this computation yielded a particularly simple result for the prefactor, namely
\beq
\texttt{simple prefactor}_{SU(2)}=1 - \frac{\gamma}{2}+\mathcal{O}(g^4) \,.
\eeq
As our very first guess we assume that our $\texttt{simple prefactor}$ in (\ref{generalForm}) takes the same form.

Next we turn to the quantum corrections to the function $\mathfrak{f}(u,v)$ in (\ref{ftree}). This function can be constrained by two simple requirements. 
\begin{itemize}
\item First we impose that our result should be invariant under $l \leftrightarrow L-l$ since this is an obvious reflection symmetry of our setup, see figure \ref{Fig3pt}. In other words, we should have 
$\mathcal{A}_{L-l}= \mathcal{A}_{l}$. 
\item Next we impose that $\mathcal{A}_{0}=0$. When $l=0$ the three point function of one non-BPS primary operator with two BPS operators formally reduces to a two point function between a non-BPS operator and a BPS operator. The latter should clearly vanish hence we impose that $\mathcal{A}_{0}=0$. 

This assumption looks very innocent but it is actually a bit trickier than it sounds; it does \textit{not} hold for the $SU(2)$ case mentioned above, for example, while \textit{naively} the exact same logic would lead to this same conclusion. For now let us ignore this subtlety; we will come back to this point in the next subsection. 
\end{itemize}
To analyze the consequence of the previous two requirements it is enough to consider the case with two excitations. According to the previous two points we should have $\mathcal{A}_{0}(u_1,u_2) =\mathcal{A}_{L}(u_1,u_2) =0$ so that  
\beqa
0&=&\mathcal{A}_{0}(u_1,u_2) \propto 1-\mathfrak f(u_1,u_2)-\mathfrak f(u_2,u_1)+1\\
0&=&\mathcal{A}_{L}(u_1,u_2) \propto 1-\mathfrak f(u_1,u_2)e^{-i L p_2}-\mathfrak f(u_2,u_1)e^{-i p_1 L}+e^{-i(p_1+p_2)L}
\eeqa
In the second line we can get rid of $L$ by using the Bethe equations $e^{-i p_1 L}= \mathcal{S}(u_1,u_2)$ and $e^{-i p_2 L}= \mathcal{S}(u_2,u_1)=1/\mathcal{S}(u_1,u_2)$. Combining both equations we find the remarkably simple result 
\beq
\mathfrak{f}(u_1,u_2)=\frac{2}{1+\mathcal{S}(u_2,u_1)} \la{Ffinal}
\eeq
which expanded in perturbation theory leads to (\ref{frakf}). Nicely, (\ref{Ffinal}) automatically leads to ${\mathcal{A}_{L-l}(u_1,\dots, u_S)}={\mathcal{A}_{l}(u_1,\dots,u_S)}$ for any $l$ and for any $S$. 

This concludes our motivation of the conjecture (\ref{mainResult}). Given that many points lack a solid derivation it is very important to check this prediction against perturbation theory to provide solid evidence for it. This was the purpose of section \ref{checksSec}.

\subsection{Further Comments and Speculative Remarks}

As anticipated above, the assumption $\mathcal{A}_0=0$ is not as innocent as it might sound since the $l\to 0$ limit might be singular. 
Indeed, this same requirement should \textit{naively} also apply to the $SU(2)$ setup studied in \cite{tailoringI} however in that case \cite{tailoringIV}\footnote{There is a slight change of notation in these expressions compared to those in \cite{tailoringI,tailoringIV}; they are related by complex conjugation. This difference comes from a different convention for the $S$-matrices used here and there. The $SU(2)$ results adapted to the conventions of the current paper are summarized in appendix \ref{su2}. }
\beqa
f_{\text{SU(2)}}(u,v) &=& \frac{u-v-i}{u-v}\left(1+\frac{g^2}{(u^2+1/4)(v^2+1/4)}+\mathcal O(g^4)\right) \nn \\
&\neq& \frac{2}{1+\mathcal{S}_\text{SU(2)}(u_2,u_1)} \nn \\ 
&=&\frac{u-v-i}{u-v}\left(1+0 \,g^2 +\mathcal O(g^4)\right)\,. \nn
\eeqa
That is, for the $SU(2)$ case the relation (\ref{Ffinal}) only holds at tree-level while in $SL(2)$ we claim that it holds at least up to one loop. 

One possible explanation is the following. When splitting an $SL(2)$ state into two subchains we get (a sum of) two decoupled Bethe states on each subchain. This holds because of the ultra-local nature of the contact terms reviewed in section \ref{ultraLocal}. This is why the $l\to 0 $ limit of the structure constant is very non-singular and our argument above for $\mathcal{A}_0=0$ should hold for our $SL(2)$ setup. 
In contradistinction, the $SU(2)$ contact terms, which appear at order $g^2$, are \textit{not} ultra local \cite{tailoringIV}. Therefore, when we cut an $SU(2)$ chain, the states on each of the two resulting subchains know about each other. As such, the $l \to 0$ limit of the $SU(2)$ structure constant is potentially more singular at one loop level. This is probably why we do not have the right to impose the $\mathcal{A}_0=0$ condition for the $SU(2)$ case beyond tree level. 

Incidentally there seems to be a nice connection between the ultra-locality of the $SL(2)$ states and some manifestation of dual conformal symmetry of the eigenstates \cite{Derkachov:2013bda,GrishaUnpublished}. Hopefully, this approach will clarify this unusual ultra-locality. Furthermore, 
if this ultra-locality were preserved at higher loops we could probably conjecture the next quantum corrections to (\ref{mainResult}) by following the logic outlined in the previous subsection.

This concludes the discussion of the subtle requirement $\mathcal{A}_0=0$. We would now like to discuss the other main requirement, namely the condition $\mathcal{A}_{L-l} = \mathcal{A}_{l}$. This relation should be quite robust since it only relies on an obvious reflection symmetry of our $SL(2)$ three point functions, see figure \ref{Fig3pt}. It is bound to work as well for the $SU(2)$ case, see figure \ref{Fig3ptSU2}.  
This condition is \textit{not} enough to fix the function $\mathfrak{f}$ completely however it \textit{does} imply that 
\beq
\mathfrak{f}(u,v) = \mathcal{S}(u,v) \mathfrak{f}(v,u) \la{fSf} \,.
\eeq
This relation is indeed satisfied both in the $SU(2)$ and in the $SL(2)$ cases at tree level and at one loop. 
\textit{If} the all loop expression for the structure constant is given by a deformation of (\ref{mainResult2}), where $\mathcal{A}_l$ is still given by a sum over partitions of Bethe roots as in (\ref{calA2}), then (\ref{fSf}) should hold to all loops. Of course, this is a big \textit{if}.

Nonetheless, the relation (\ref{fSf}) resembles some sort of Watson equation for form factors \cite{watson}. 
Perhaps this is more than a coincidence. After all, as recently advocated \cite{tristan}, it is natural to expect form factors to play an important role in the study of three-point functions. Can $\mathfrak{f}$ be given some nice finite coupling definition in terms of form factors of the BMN string? If so, we might hope to bootstrap it exactly.

Relations of the form (\ref{fSf}) recently played a central role in a very different context, namely in the computation of null polygonal Wilson loops in planar $\mathcal{N}=4$ SYM theory \cite{short}. There, the so called \textit{fundamental relation} is a functional equation for the so called pentagon transitions $P$ which reads \cite{short}
\beq
P(u|v)=S(u,v) P(v|v)
\eeq 
where ${S}(u,v)$ is the S-matrix for the fundamental excitations \textit{on top of the GKP state}. One might wildly speculate whether the pentagon transitions $P(u|v)$ and our function $\mathfrak{f}(u,v)$ are not so different after all. Can it be that they are similarly defined objects but naturally defined on top of different \textit{vacua}? (namely the BMN string \cite{BMN} for $\mathfrak{f}(u,v)$ and the GKP string \cite{GKP} for $P(u|v)$) 

It would be very interesting to study the next loop correction to (\ref{mainResult2}) and confirm that it is still given by some simple deformation of (\ref{calA2}) and (\ref{calB2}). This would definitely give very strong support to these speculative ideas and strongly motivate us to push them further. 

Along these lines, it would be interesting to investigate whether the strong coupling results \cite{strongCouplingPapers} can be written as (the classical continuum limit of) some strong coupling deformation of (\ref{calA2}) or of its scalar counterpart \cite{tailoringIII,tailoringIV}.  

Finally, it would be fascinating to develop further the (algebraic) integrabity description of the eigenstates of the dilatation operator at higher loops. For operators with derivatives we have roughly no control over the operators beyond one loop. With an algebraic description \`a la \cite{tailoringIV,didina1,didina2,niklasLong} we would be able to make substantially more \textit{rigorous} progress, with considerably less guesswork.

 \section*{Acknowledgements}
We thank Fernando Alday, Gleb Aryutunov, Benjamin Basso, Joao Caetano, 
Dima Chicherin, Thiago Fleury, Vasco Goncalves, Kolya Gromov, Yunfeng Jiang, Volodya Kazakov, Grisha Korchemsky, Joao Penedones, Evgeny Sobko, Jonathan Toledo, Amit Sever and Emery Sokatchev for numerous discussions and correspondence. We are grateful to Grisha Korchemsky for exchanges on the ultra-local property of the quantum corrected wave functions. We thank Dima Chicherin and Emery Sokatchev for sharing with us their unpublished results \cite{emeryToAppear}. We thank Nikolay Gromov, Yunfeng Jiang and Amit Sever for related collaborations (on this and closely related topics). We thank the ICTP-SAIFR for warm hospitality during the completion stages of this project. 
 Research at the Perimeter Institute is supported in part by the Government of Canada through NSERC and by the Province of Ontario through MRI.

\appendix

\section{More details on the twist $4$ analysis}\la{detailsT4}
In \cite{Arutyunov:2003ad,Dolan:2004iy} the general four-point function for the following scalar operators is given explicitly up to two loops,
\begin{eqnarray}
\mathcal O^{(p)}_j (x_j, t_j)= t^{a_1}_j\cdots t^{a_p}_j \text{Tr}\left(\phi_{a_1}\cdots\phi_{a_p}\right)(x_j)\,, 
\end{eqnarray}
where $p$ is the twist of the BPS operators and $j=1,\dots, 4$. The polarization vector $t_j$ is null and the index $a_n=1,\dots, 6$ in the usual $SO(6)$ R-charge index. To make contact with our correlation function (\ref{4pt}) we choose $p=3$ and take the polarization vectors to be the following:
\begin{eqnarray}
&&t_1 = \left( 1,i, i\alpha_1, -\alpha_1,0,0\right) \,\,\,\,\,\, , \qquad  t_2 = \left( 1,i, i\alpha_2,\alpha_2,0,0\right) \, ,\\
&&t_3 = \left( 1,-i,0,0,i\alpha_3,-\alpha_3\right) \, , \qquad 
t_4 = \left( 1,-i,0,0,i\alpha_4,\alpha_4\right) \,. 
\end{eqnarray}
We adopt the usual notation $Z=\phi_1+i\phi_2, X=\phi_3+i\phi_4, Y=\phi_5+i\phi_6$ and similar complex conjugate expressions for $\bar Z,\bar X,\bar Y$. Now it is obvious that
\begin{eqnarray}
\text{Tr}(ZZX) =\[ \frac{1}{3 i}\frac{\partial}{\partial\alpha_1}\mathcal O^{(3)}_1 \]_{\alpha_1=0}\,,
\end{eqnarray}
and likewise for the other external operators in (\ref{4pt}). Therefore the polarized four-point function (\ref{4pt}) is obtained by
\begin{eqnarray}
\frac{1}{x^6_{12} x^6_{34}}\mathcal G(z,\bar z) = \frac{1}{3^4} \Big[\prod\limits^{4}_{j=1}\frac{\partial}{\partial\alpha_j}\<\prod\limits^4_{j=1}\mathcal O^{(3)}_j(x_j,t_j)\>\Big]_{\alpha_j=0}\,.
\end{eqnarray}
In this way we read off $\mathcal G(z,\bar z)$ from \cite{Arutyunov:2003ad,Dolan:2004iy}. Its OPE expansion is given in (\ref{pertvoila}). 

\section{Explicit expressions for the sums $\mathcal{P}_S^{(n,m)}$} \la{explicitSums}
In perturbation theory we have 
\begin{eqnarray}
\Delta_{\mathbf u} &=& L+S+g^2 \gamma^{(1)}_{\mathbf u}+g^4 \gamma^{(2)}_{\mathbf u} +g^6 \gamma^{(3)}_{\mathbf u}+\cdots \,,\nonumber\\
C^{\bullet\circ\circ}_{\mathbf u} &=& C^{(0)}_{\mathbf u}+g^2 C^{(1)}_{\mathbf u}+ g^4 C^{(2)}_{\mathbf u}+g^6 C^{(3)}_{\mathbf u}+\cdots \, , \nonumber
\end{eqnarray}
such that, from (\ref{defSums}) we find (the sums are sums over all solutions to Bethe equations for a given $L$ and $S$)
\begin{eqnarray}
\mathcal P_{S}^{(0,0)}&=& \sum_{\{u_j\}} \left(C^{(0)}_{\mathbf u}\right)^2\, ,\la{p0} \\
\mathcal P_{S}^{(1,0)}&=& \sum_{\{u_j\}} 2 C^{(0)}_{\mathbf u} C^{(1)}_{\mathbf u}\, ,\la{p10} \\
\mathcal P_{S}^{(1,1)}&=& \sum_{\{u_j\}} \gamma^{(1)}_{\mathbf u}\left(C^{(0)}_{\mathbf u}\right)^2\, ,\la{p11} \\
\mathcal P_{S}^{(2,0)}&=&\sum_{\{u_j\}}\Big[\left(C^{(1)}_{\mathbf u}\right)^2+2C^{(0)}_{\mathbf u}C^{(2)}_{\mathbf u}\Big]\, ,\la{p20}\\
\mathcal P_{S}^{(2,1)}&=&\sum_{\{u_j\}}\Big[2\gamma^{(1)}_{\mathbf u}C^{(1)}_{\mathbf u}C^{(0)}_{\mathbf u}+\gamma^{(2)}\left(C^{(0)}_{\mathbf u}\right)^2\Big]\, ,\la{p21}\\
\mathcal P_{S}^{(2,2)}&=&\sum_{\{u_j\}}\frac{1}{2} \left(\gamma^{(1)}_{\mathbf u}\right)^2\left(C^{(0)}_{\mathbf u}\right)^2\, , \la{p22} \\
\mathcal P_{S}^{(3,0)} &=& \sum_{\{u_j\}} 2\left(C^{(0)}_{\mathbf u} C^{(3)}_{\mathbf u}+C^{(1)}_{\mathbf u} C^{(2)}_{\mathbf u}\right)\,, \la{p30} \\
\mathcal P_{S}^{(3,1)} &=& \sum_{\{u_j\}} \Big[\gamma^{(1)}\Big(\left(C^{(1)}_{\mathbf u}\right)^2+2C^{(0)}_{\mathbf u} C^{(2)}_{\mathbf u}\Big)+2\gamma^{(2)}C^{(0)}_{\mathbf u}C^{(1)}_{\mathbf u}+\gamma^{(3)} \left(C^{(0)}_{\mathbf u}\right)^2\Big] \,, \la{p31} \\
\mathcal P_{S}^{(3,2)}&=& \sum_{\{u_j\}}\Big[\left(\gamma^{(1)}_{\mathbf u}\right)^2C^{(1)}_{\mathbf u} C^{(0)}_{\mathbf u}+\gamma^{(1)}_{\mathbf u} \gamma^{(2)}_{\mathbf u}\left(C^{(0)}_{\mathbf u}\right)^2\Big]\,,\\
\mathcal P_{S}^{(3,3)}&=& \sum_{\{u_j\}}\frac{1}{6} \left(\gamma^{(1)}_{\mathbf u}\right)^3\left(C^{(0)}_{\mathbf u}\right)^2\,,\la{p33}
\end{eqnarray} 
etcetera.

\section{Other examples: Twist $3$, Twist $6$ and a special Twist $4$} \la{twist3}

In this appendix, we present some more sum rules and their relation to the conformal partial wave expansion of the corresponding four-point functions. 
\subsection{Sums from Integrability and Comparison with One Loop Correlators}
Here we present the sums (\ref{defSums}) and (\ref{defSumsNew}) obtained by solving the Bethe equations and using (\ref{mainResult}) for a few different cases, see tables \ref{PnmTwist3}, \ref{PnmTwist6} and \ref{tildePnm}.

 \begin{table}[h!]
\renewcommand{\arraystretch}{1.5}
\caption{Values of $\mathcal P^{(n,m)}_S$ for $L=3$, $l=2$ and $N=1$.}\label{PnmTwist3}
\centering
\begin{tabular}{|c|ccccccc|}
 \firsthline
 Spin-$S$ & $0$ & $2$ & $3$ & $4$ & $5$ & $6$ & $\dots$ \\ \hline
 $\mathcal P^{(0,0)}$ & $3$ & $\frac{1}{2}$ & $ \frac{2}{35}$ &  $\frac{1}{18}$ & $\frac{ 2}{231}$ & $\frac{3}{572}$ & $\dots$ \\ \hline
 $\mathcal P^{(1,1)}$ & $0$ & $4$ & $ \frac{6}{7}$ &  $\frac{2}{3}$ & $\frac{ 5}{33}$ & $\frac{ 166}{2145}$ & $\dots$ \\ \hline
 $\mathcal P^{(1,0)}$ & $0$ & $-4$ & $ -\frac{41}{49}$ &  $-\frac{35}{54} $ & $ -\frac{1907}{13068} $ & $ -\frac{681653}{9202050} $ & $\dots$ \\ \hline
 $\mathcal P^{(2,2)}$ & $0$ & $16$ &  $\frac{45}{7}$ & $ 4$ & $ \frac{175}{132} $ & $\frac{ 73721}{128700} $ & $\dots$ \\ \hline
 $\mathcal P^{(2,1)}$ & $0$ & $-44$ & $ -\frac{1545}{98} $ & $ -\frac{179}{18} $ & $ -\frac{163135}{52272} $ & $ -\frac{1499665543}{1104246000} $ & $\dots$ \\ \hline
 $\mathcal P^{(3,3)}$ & $0$ & $\frac{128}{3} $ & $\frac{ 225}{7}$ &  $16$ & $\frac{ 6125}{792}$ & $\frac{ 32901001}{11583000} $ & $\dots $ \\ \hline
 $\mathcal P^{(3,2)}$ & $0$ & $-224$ & $-\frac{6975}{49} $ & $ -\frac{218}{3}$ & $ -\frac{62475}{1936} $ & $ -\frac{199401651971}{16563690000} $ & $\dots$ \\ \hline
\lasthline
\end{tabular}
\end{table}

\begin{table}[h!]
\renewcommand{\arraystretch}{1.5}
\caption{Values of $\mathcal P^{(n,m)}_S$ for $L=6$, $l=3$ and $N=1$.}\label{PnmTwist6}
\centering
\begin{tabular}{|c|cccccc|}
 \firsthline
Spin-$S$ & $0$ &$ 2$ & $4$  &$6$ & $8$  &$\dots $ \\ \hline
$\mathcal P^{(0,0)}$ & $6$ & $\frac{26}{7}$  & $\frac{12}{11}$ & $\frac{12}{55}$ & $\frac{145}{4199}$ & $\dots$ \\ \hline
$\mathcal P^{(1,1)}$ & $0$ & $ 12 $  &  $ \frac{76}{11}$ &  $\frac{124}{65}$ &  $\frac{831}{2261}$ & $\dots$ \\ \hline
$\mathcal P^{(1,0)}$ & $0$ & $-12 $ &  $-\frac{2494}{363} $ &  $-\frac{15917}{8450}$ &  $-\frac{9212823}{25560605}$ & $\dots$ \\ \hline
$\mathcal P^{(2,2)}$ & $0$ & $24$ & $\frac{2666}{99}$  &  $\frac{85441}{8775}$ &  $\frac{7855283}{3561075}$ & $\dots$ \\ \hline
$\mathcal P^{(2,1)}$ & $0$ & $-64$ & $-\frac{73598}{1089}$ &  $-\frac{162099173}{6844500}$ & $ -\frac{5071321422691}{966190869000}$ & $\dots$ \\ \hline
$\mathcal P^{(3,3)}$ & $0$ & $\frac{160}{3}$ & $ \frac{80336}{891}$ & $ \frac{30768139}{789750} $ & $ \frac{2972169757}{299130300}$ & $\dots$ \\ \hline
$\mathcal P^{(3,2)}$ & $0$ & $-256$ & $ -\frac{3998960}{9801}$ &  $ -\frac{17335124569}{102667500}$  & $
-\frac{37633529965763}{901778144400}$ & $\dots$ \\ \hline
\lasthline
 \end{tabular}
 \end{table}

\begin{table}[!h]
\renewcommand{\arraystretch}{1.5}
\caption{Values of $\widetilde{\mathcal{P}}^{(n,m)}_S$ (see (\ref{defSumsNew}) in section \ref{twist4Main})}\label{tildePnm}
\centering
\begin{tabular}{|c|ccccccc|}
 \firsthline
Spin-$S$ & $0$ &$ 2$ & $4$  &$6$ & $8$ & $10$ &$\dots $ \\ \hline
$\widetilde{\mathcal{P}}^{(0,0)}$ & $4$ & $\frac{4}{5}$ & $ \frac{2}{21}$ &  $\frac{4}{429}$ & $\frac{2}{2431}$ & $ \frac{2}{29393}$ & $\dots$ \\ \hline
$\widetilde{\mathcal{P}}^{(1,1)}$ & $0$ &  $4$ &  $\frac{2}{3}$ &  $\frac{166}{2145}$ & $ \frac{59}{7735}$ & $ \frac{487}{712215}$ & $\dots$ \\ \hline
$\widetilde{\mathcal{P}}^{(1,0)}$ & $0$ & $-4$ &  $-\frac{35}{54}$ & $-\frac{681653}{9202050}$ & $ -\frac{7775819}{1076944050}$ & $
-\frac{185976967}{289857260700}$ & $\dots$ \\ \hline
$\widetilde{\mathcal{P}}^{(2,2)}$ & $0$ & $8$ & $\frac{14}{9}$ & $\frac{6421}{32175}$ & $\frac{34361}{1624350}$ & $\frac{ 450187}{224347725} $ & $\dots$ \\ \hline
$\widetilde{\mathcal{P}}^{(2,1)}$ & $0$ &  $-24$ &  $-\frac{118}{27}$ & $ -\frac{1376526}{2556125}$ & $ -\frac{12543017063}{226158250500} $ &  
$-\frac{313211878949}{60870024747000} $ & $\dots$ \\ \hline
$\widetilde{\mathcal{P}}^{(3,3)}$ & $0$ &  $0$ &  $-\frac{232}{81} $ & $ -\frac{54107}{87750}$ &  $-\frac{5029138}{59038875}$ & $
-\frac{3249557287}{339213760200}$ & $\dots$ \\ \hline
$\widetilde{\mathcal{P}}^{(3,2)}$ & $0$ & $ -24$ & $ \frac{931}{162}$ & $\frac{ 131592343}{83655000} $ &  $\frac{1119943628459}{4871100780000}$ & $
\frac{1736255596975721}{65739626726760000}$ & $\dots$ \\ \hline
\lasthline
 \end{tabular}
 \end{table}

These sums can be checked against the OPE expansion of four-point correlation functions. Indeed, table \ref{PnmTwist3} governs the leading twist behaviour of 
\begin{eqnarray}
\< \text{Tr}(ZX)\!(x_1)\, \text{Tr}(ZZ\bar X)\!(x_2) \, \text{Tr}(\bar ZY)\!(x_3)\, \text{Tr}(\bar Z\bar Z\bar Y)\!(x_4) \> = \frac{1}{x^5_{12} x^5_{34}} \frac{|x_{13}|}{|x_{24}|} \mathcal G(z,\bar z)\,, \la{4ptT3}
\end{eqnarray}
while table \ref{PnmTwist6} governs the leading twist behaviour of 
\begin{eqnarray}
\< \text{Tr}(ZZZX)\!(x_1)\, \text{Tr}(ZZZ\bar X)\!(x_2) \, \text{Tr}(\bar Z\bar Z\bar ZY)\!(x_3)\, \text{Tr}(\bar Z\bar Z\bar Z\bar Y)\!(x_4) \> = \frac{1}{x^6_{12} x^6_{34}}\mathcal G(z,\bar z)\,. \la{4ptT6}
\end{eqnarray}
The comparison between the OPE decomposition of (\ref{4ptT3}) and (\ref{4ptT6}) and the corresponding tables \ref{PnmTwist3} and \ref{PnmTwist6} goes exactly as for the example discussed in section \ref{twist4Main}. Here, we extracted the correlation functions (\ref{4ptT3}) and (\ref{4ptT6}) from \cite{Berdichevsky:2007xd,D'Alessandro:2005dq} and \cite{Arutyunov:2002fh} respectively. 

As for table \ref{tildePnm}, it controls the leading twist behaviour of the asymmetric correlator (\ref{4ptNew}). It would be very interesting to perform a perturbative computation of this object and check whether it matches the Integrability predictions. According to the latter, we should have 
$\tilde{\mathcal G}(z,\bar z) = \bar z^{2} \tilde f(z,\tau) + \mathcal{O}(\bar z^3)$ where $\tilde f(z,\tau)$ reads
  \begin{eqnarray}
\tilde f(z,\tau) &=& g^0 \left[4 z^2 + 6 z^3 + 8 z^4 + 10 z^5 + 12 z^6+\cdots\right] \nonumber\\
&+& g^2 \left[\tau\left(4 z^4 + 10 z^5 + \frac{52}{3} z^6+\cdots\right)-\left(4 z^4 + 9 z^5 + \frac{130}{9} z^6+\cdots\right)\right]\nonumber\\
&+& g^4 \left[\tau^2\left(8 z^4 + 20 z^5 + \frac{314}{9} z^6+\cdots\right)-\tau\left(24 z^4 + 56 z^5 + \frac{836}{9} z^6+\cdots\right)+\mathcal O(\tau^0)\right]\nonumber\\
&+& g^6 \left[\tau^3 \left(-\frac{232}{81} z^6+\cdots\right) -\tau^2\left(24 z^4 + 60 z^5 + \frac{15269}{162} z^6+\cdots\right)+\mathcal O(\tau)\right]\nonumber\\
&+& \mathcal O(g^8)\,. \la{f2433}
\end{eqnarray}

\subsection{Two Loop Sums from 4pt Correlation Functions}
The correlators (\ref{4ptT3}) and (\ref{4ptT6}) were actually computed up to 2-loop order. Hence, from their OPE analysis, we can extract new predictions for the sums $\mathcal P^{(2,0)}_S$, see tables \ref{P20Twist3} and \ref{P20Twist6}. As discussed in the main text, these predictions can hopefully be used to constrain the quantum corrections to (\ref{mainResult}). 

\begin{table}[h!]
\renewcommand{\arraystretch}{1.5}
\caption{Values of $\mathcal P^{(2,0)}_S$ for $L=3$, $l=2$ and $N=1$.}\label{P20Twist3}
\centering
\begin{tabular}{|c|ccccccc|}
 \firsthline
 Spin-$S$ & $0$ & $2$ & $3$ & $4$ & $5$ & $6$ & $\dots$ \\ \hline
$\mathcal P^{(2,0)}$ & $0$ & $40 + 24 \zeta_3 $ &  $\frac{17251}{1372} + \frac{36}{7}\zeta_3 $ & $\frac{ 698}{81} + 
 4 \zeta_3 $ & $ \frac{51784253}{20699712} + \frac{
 10}{11} \zeta_3 $ & $\frac{ 10948488366143}{9474430680000} + \frac{332 }{715}\zeta_3 $ & $ \dots $ \\
 \lasthline
 \end{tabular}
 \end{table}

\begin{table}[h!]
\renewcommand{\arraystretch}{1.5}
\caption{Values of $\mathcal P^{(2,0)}_S$ for $L=6$, $l=3$ and $N=1$.}\label{P20Twist6}
\centering
\begin{tabular}{|c|cccccc|}
 \firsthline
Spin-$S$ & $0$ &$ 2$ & $4$  &$6$ & $8$  &$\dots $ \\ \hline  
$\mathcal P^{(2,0)}$ & $0$ & $56 + 72 \zeta_3$ & $ \frac{2041118}{35937} + \frac{
 456}{11} \zeta_3$ & $ \frac{6550035818}{333669375} + \frac{
 744}{65} \zeta_3$   & $\frac{188739630680649539}{43691151096180000} + \frac{4986}{2261} \zeta_3$& $\dots$ \\ \hline
\lasthline
\end{tabular}
\end{table}

\section{Two- and three-magnon contact terms}\la{ContTerm}
\subsection{Contact terms} \la{D1}
The contact terms $\mathbb{C}_{\substack{\bullet\vspace{-.6mm}\\ \bullet}}(p_a,p_b) $ and $\mathbb{C}_{\substack{\bullet\vspace{-.6 mm}\\ \bullet\vspace{-.6 mm} \\ \bullet}}(p_a,p_b,p_c) $ discussed in section \ref{ultraLocal} can be found by direct diagonalization of the two loop dilatation operator of Eden and Staudacher \cite{Eden:2006rx}. For two particles \cite{Eden:2006rx} 
\beq
\mathbb{C}_{\substack{\bullet\vspace{-.6mm}\\ \bullet}}(p_1,p_2)=2\left(\sin^2\frac{p_1}{2}+\sin^2\frac{p_2}{2}-\frac{1}{2}\sin^2\frac{p_1+p_2}{2}\right)\,, \la{ES1}
\eeq
while for three particles we find 
\beq
\mathbb{C}_{\substack{\bullet\vspace{-.6 mm}\\ \bullet\vspace{-.6 mm} \\ \bullet}}(p_1,p_2,p_3)=\frac{8}{3}\(\sin^2 \frac{p_1}{2}+\sin^2 \frac{p_2}{2}+\sin^2 \frac{p_3}{2}-\frac{1}{3}\sin^2 \frac{p_1+p_2+p_3}{2}\)\,. \la{ESC3}
\eeq 
Someone bold could be tempted to conjecture that $\mathbb{C}_{\substack{\bullet\vspace{-.2cm}\\ \vdots \vspace{-.4 mm} \\ \bullet}}(p_1,\dots,p_N)\propto  \sin^2\frac{p_1}{2}+\dots + \sin^2\frac{p_N}{2}-\frac{1}{N} \sin^2\frac{p_1+\dots+p_N}{2}$; unfortunately we have some preliminary evidence against this natural guess.

\subsection{Zwiebel and Eden-Staudacher representations}\la{D2}
As mentioned in section \ref{ultraLocal} another two dilation operator was proposed by Zwiebel in \cite{Zwiebel:2008gr}. It should differ from the one in \cite{Eden:2006rx} by a similarity transformation, i.e. by a change of basis. For example, for $2$ and $3$ particles we found that the two wave functions are simply related by a change of basis given by 
\begin{eqnarray}
\psi^{\text{Z}}_{n,n} &=& \psi^{\text{ES}}_{n,n}+\frac{1}{4}g^2\Big(\psi^{\text{ES}}_{n-1,n}+\psi_{n,n+1}^{\text{ES}}\Big) \, ,\nonumber\\
\psi^{\text{Z}}_{n,n+1} &=& \psi_{n,n+1}^{\text{ES}}-\frac{1}{4}g^2\Big(\psi_{n,n}^{\text{ES}}+\psi_{n+1,n+1}^{\text{ES}}\Big)\, ,  \nn 
\end{eqnarray}
and 
\begin{eqnarray}
\psi^{\text{Z}}_{n,n,n} &=& \psi_{n,n,n}^{\text{ES}}+\frac{g^2}{3}\left(\psi_{n,n,n+1}^{\text{ES}}+\psi_{n-1,n,n}^{\text{ES}}\right)+\frac{g^2}{6}\left(\psi_{n,n+1,n+1}^{\text{ES}}+\psi_{n-1,n-1,n}^{\text{ES}}\right),\nonumber\\
\psi^{\text{Z}}_{n,n,n+1} &=& \psi_{n,n,n+1}^{\text{ES}}-\frac{g^2}{3}\psi_{n,n,n}^{\text{ES}}+\frac{g^2}{4}\psi_{n-1,n,n+1}^{\text{ES}}-\frac{g^2}{6}\psi_{n,n+1,n+1}^{\text{ES}},\nonumber\\
\psi^{\text{Z}}_{n-1,n,n} &=& \psi_{n-1,n,n}^{\text{ES}}-\frac{g^2}{3}\psi_{n,n,n}^{\text{ES}}+\frac{g^2}{4}\psi_{n-1,n,n+1}^{\text{ES}}-\frac{g^2}{6}\psi_{n-1,n-1,n}^{\text{ES}},\nonumber\\
\psi^{\text{Z}}_{n-1,n,n+1} &=& \psi_{n-1,n,n+1}^{\text{ES}}-\frac{g^2}{4}\left(\psi_{n,n,n+1}^{\text{ES}}+\psi_{n-1,n,n}^{\text{ES}}+\psi_{n-1,n-1,n+1}^{\text{ES}}+\psi_{n-1,n+1,n+1}^{\text{ES}}\right).\nonumber
\end{eqnarray}
while for excitations which are more widely separated there is no need for any change of basis. 
In the Zwiebel's basis the wave functions still take the form (\ref{waveFunctions}) but now all of the contact terms in (\ref{C2}) and (\ref{C3}) are present. In other words, the nice ultra-local nature of the contact terms alluded to in section \ref{ultraLocal} is lost. For example, 
\begin{eqnarray}
 \mathbb{C}^{Z}_{\bullet\bullet}(p_1,p_2) &=& -\frac{\cos\left(\frac{p_1}{2}+\frac{p_2}{2}\right)}{2 \cos\frac{p_1}{2}\cos\frac{p_2}{2}+6 \sin\frac{p_1}{2}\sin\frac{p_2}{2}}\, , \la{Z1}\\
 \mathbb{C}^{Z}_{\substack{\bullet\vspace{-.6 mm} \\ \bullet}}(p_1,p_2) &=& \sin^2\frac{p_1}{2}+\sin^2\frac{p_2}{2}-\frac{1}{2}\sin^2\frac{p_1+p_2}{2}+\frac{1}{2}\, . \la{Z2}
\end{eqnarray}
The three particle contact terms can also be trivially obtained by following the change of basis written above but they are quite messy and unilluminating to be written down.  

\section{Comparison with $SU(2)$}\la{su2}

\begin{figure}[t]
\centering
\includegraphics[scale=0.5]{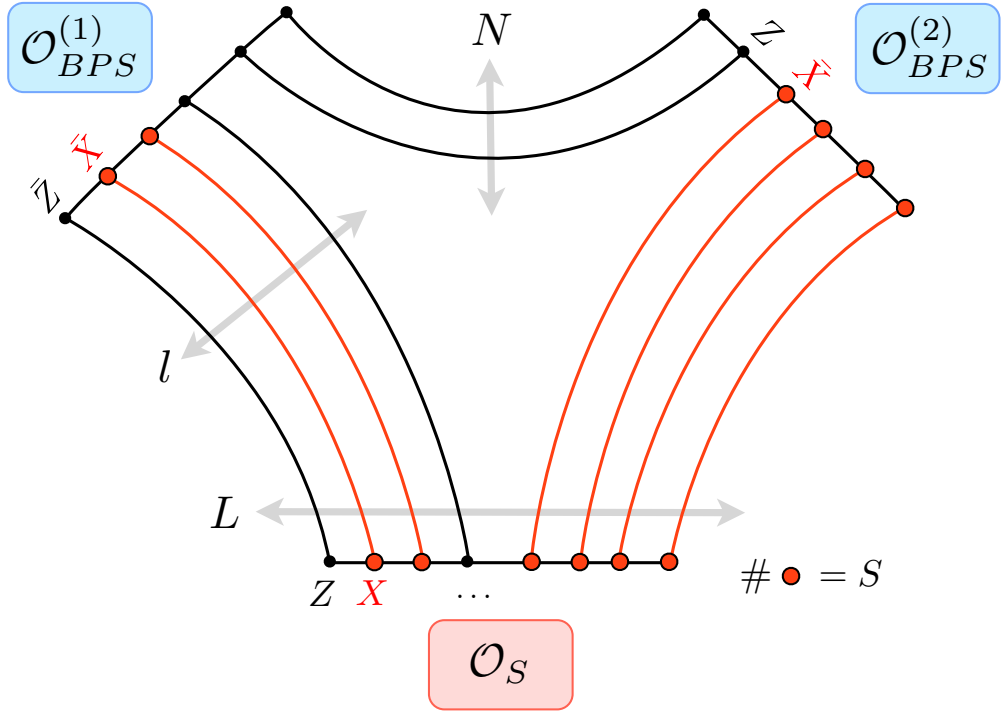}\vspace{-.5cm}
\caption{In \cite{tailoringI} and \cite{tailoringIV} the structure constant involving three $SU(2)$ operators as depicted in this figure was computed at tree level and at one loop respectively. 
} \label{Fig3ptSU2} 
\end{figure}

Here we review the value of the $SU(2)$ structure constant -- depicted in figure \ref{Fig3ptSU2} -- at tree level and one loop \cite{tailoringIV}. We massage slightly the result in that paper by conjugating it and then choosing the phase of the non-BPS operator so that the structure constant is real (for real roots). We have 
\begin{eqnarray}
C^{\bullet\circ\circ}_{\text{SU(2)}} = \frac{\sqrt{L(\sc l+N)(L-\sc l+N)}}{\sqrt{
\left(\!\!\begin{array}{c}\sc l+N\\N\end{array}\!\!\right)\left(\!\!\begin{array}{c}L-\sc l+N\\N\end{array}\!\!\right)}} \left(1-\frac{\gamma }{2}\right)\frac{ \mathcal A_{\sc l}^{\text{SU(2)}}  }{\mathcal B_{\text{SU(2)}} }\,+O(g^4) \,.
\end{eqnarray}
where
\begin{eqnarray}
\mathcal B_{\text{SU(2)}} =\sqrt{ {\frac{1}{\prod\limits_{j=1}^S \frac{\partial}{\partial u_j}\left[\frac{1}{i}\log\frac{x^+_j}{x^-_j}\right]} \det_{1\le j,k \le S}\frac{\partial}{\partial u_j}\left[iL \log\frac{x^+_k}{x^-_k}+i\sum_{l\neq k}^S\log \mathcal S_{\text{SU(2)}}(u_k, u_l)\right]}}\, . \la{Bsu2}
\end{eqnarray}
with the $SU(2)$ S-matrix 
\begin{eqnarray}
\mathcal{S}_{\text{SU(2)}}(u,v) = \frac{u-v-i}{u-v+i}\,.
\end{eqnarray}
Finally
\begin{eqnarray}
\mathcal A_{\sc l}^{\text{SU(2)}} = \frac{\sum\limits_{{\color{red}\alpha}\cup{\color{blue} \bar\alpha} = \{u_j\}} (-1)^{|{\color{red} \alpha}|} \prod\limits_{\color{blue} u_j\in \bar\alpha} \left[\frac{x^+({\color{blue} u_j})}{x^-({\color{blue} u_j})}\right]^{-\sc l}\prod\limits_{\substack{{\color{red} u_j\in\alpha}\\ {\color{blue} u_k\in\bar\alpha}}} \mathfrak f_{\text{SU(2)}}({\color{red} u_j},{\color{blue} u_k})}{\sqrt{\prod\limits_{j\neq k}\mathfrak f_{\text{SU(2)}}(u_j,u_k)}\prod\limits_{j} \(1-\frac{x^-_j}{x_j^+}\)}\,,\la{su2calA}
\end{eqnarray}
where 
\begin{eqnarray}
\mathfrak f_{\text{SU(2)}}(u,v) = \frac{u-v-i}{u-v}\left(1+\frac{g^2}{(u^2+1/4)(v^2+1/4)}+\mathcal O(g^4)\right)\,.
\end{eqnarray}
The reader might be puzzled as the S-matrix $\mathcal{S}_{\text{SU(2)}}(u,v) $ and the function $\mathfrak f_{\text{SU(2)}}(u,v)$ in this appendix are the complex conjugate of the expressions reported in \cite{tailoringIV}. However, we should note that the $i p(u_j)L$ factors in (\ref{Bsu2}) and (\ref{su2calA}) also appear with an opposite sign compared with \cite{tailoringIV}. That is, our expressions are simply the complex conjugate of those in \cite{tailoringIV}. Since the final result is real this complex conjugation is not an issue. We chose to perform this conjugation to highlight the similarities between the $SU(2)$ and the $SL(2)$ results in the conventions (for the S-matrix) used in this paper.

\end{document}